\documentclass[fleqn,usenatbib]{mnras}

\usepackage[T1]{fontenc}
\usepackage{ae,aecompl}
\usepackage[utf8]{inputenc}
\usepackage{graphicx}	% Including figure files
\usepackage{amsmath}	% Advanced maths commands
\usepackage{amssymb}	% Extra maths symbols
\usepackage{float}
\usepackage[normalem]{ulem}
\usepackage{booktabs}	
\usepackage{hyperref}

\hypersetup{draft}

\title[Effect of MSSFR on DCO mergers]{
The effect of the metallicity-specific star formation history on double compact object mergers}

\author[C. J. Neijssel et al.]{\parbox{\textwidth}{
Coenraad J. Neijssel,$^{1,7}$ \thanks{E-mail: cneijssel@star.sr.bham.ac.uk}
Alejandro Vigna-G\'{o}mez,$^{1,5,6}$
Simon Stevenson,$^{2,6}$
Jim W. Barrett,$^{1,9}$
Sebastian M. Gaebel,$^{1}$
Floor Broekgaarden,$^{3,4,5,6}$
Selma E. de Mink,$^{3,4}$
Dorottya Sz\'{e}csi,$^{1,10}$
Serena Vinciguerra$^{1,7,8}$
and Ilya Mandel$^{5,6,1}$ 
\\ 
}
\vspace{0.5cm}\\
\parbox{\textwidth}{
$^{1}$Birmingham Institute for Gravitational Wave Astronomy and School of Physics and Astronomy,\\
University of Birmingham, Birmingham, B15 2TT, United Kingdom\\
$^{2}$Centre for astrophysics and supercomputing, Swinburne University of Technology, Hawthorn VIC 3122, Australia \\
$^{3}$Anton Pannekoek Institute for Astronomy, University of Amsterdam, Postbus 94249, 1090 GE Amsterdam, The Netherlands \\
$^{4}$GRAPPA, University of Amsterdam, Science Park 904, 1098 XH Amsterdam, The
Netherlands \\
$^{5}$Monash Centre for Astrophysics, School of Physics and Astronomy, Monash University, Clayton, Victoria 3800, Australia\\
$^{6}$The ARC Center of Excellence for Gravitational Wave Discovery -- OzGrav\\
$^{7}$Albert-Einstein-Institut, Max-Planck-Institut f\"{u}r Gravitationsphysik, D-30167 Hannover, Germany \\
$^{8}$Leibniz Universit\"{a}t Hannover, D-30167 Hannover, Germany\\
$^{9}$Klarna Bank AB (publ). Sveav\"{a}gen 46, 111 34 Stockholm\\
$^{10}$I. Physikalisches Institut, Universit\"{a}t zu K\"{o}ln, Z\"{u}lpicher-Strasse 77, D-50937 Cologne, Germany\\
}
}

\date{Accepted XXX. Received YYY; in original form ZZZ}

\pubyear{xxxx}

\usepackage{color}
\usepackage{pdflscape} %based on guideline https://mirror.hmc.edu/ctan/macros/latex/contrib/mnras/mnras_guide.pdf
\usepackage{lscape}
\usepackage{rotating}
\usepackage{acronym}   

\definecolor{ochre}{rgb}{0.8, 0.47, 0.13}
\acrodef{NS}{neutron star}
\acrodef{BH}{black hole}
\acrodef{BBH}{binary black hole}
\acrodef{BNS}{binary neutron star}
\acrodef{BHNS}{black hole -- neutron star binary}

\acrodef{ZAMS}{zero-age main sequence}
\acrodef{MS}{main sequence}
\acrodef{HG}{Hertzsprung gap}
\acrodef{aLIGO}{advanced Laser Interferometer Gravitational-wave Observatory}

\acrodef{DCO}{double compact object}
\acrodef{IMF}{initial mass function}
\acrodef{LBV}{luminous blue variable}
\acrodef{SN}{supernova}
\acrodefplural{SN}[SNe]{supernovae}
\acrodef{PISN}{pair-instability supernova}
\acrodefplural{PISN}[PISNe]{pair-instability supernovae}
\acrodef{CE}{common-envelope}
\acrodef{RLOF}{Roche-lobe overflow}

\acrodef{SFR}{star formation rate}
\acrodef{MSSFR}{metallicity-specific star formation rate}
\acrodef{GSMF}{galaxy stellar-mass function}
\acrodef{MZ}{galaxy stellar mass -- metallicity}
\acrodef{PDF}{probability density function}
\acrodef{SNR}{signal to noise ratio}
\acrodef{SSE}{single stellar evolution}

\begin{document}
\label{firstpage}
\pagerange{\pageref{firstpage}--\pageref{lastpage}}
\maketitle

\begin{abstract}
We investigate the impact of uncertainty in the metallicity-specific star formation rate over cosmic time on predictions of the rates and masses of double compact object mergers observable through gravitational waves.  We find that this uncertainty can change the predicted detectable merger rate by more than an order of magnitude, comparable to contributions from uncertain physical assumptions regarding binary evolution, such as mass transfer efficiency or supernova kicks.  We statistically compare the results produced by the COMPAS population synthesis suite against a catalog of gravitational-wave detections from the first two Advanced LIGO and Virgo observing runs.   We find that the rate and chirp mass of observed binary black hole mergers can be well matched under our default evolutionary model with a star formation metallicity spread of $0.39$ dex around a mean metallicity $\left<Z\right>$  that scales with redshift $z$ as $\left<Z\right>=0.035 \times 10^{-0.23 z}$, assuming a star formation rate of $0.01 \times (1+z)^{2.77} / (1+((1+z)/2.9)^{4.7}) \, \rm{M}_\odot$  Mpc$^{-3}$ yr$^{-1}$. 
Intriguingly, this default model predicts that 80\% of the approximately one binary black hole merger  per day that will be detectable at design sensitivity will have formed through isolated binary evolution with only dynamically stable mass transfer, i.e., without experiencing a common-envelope event.

\end{abstract}


%%%%%%%%%%%%%%%%%%%%%%%%%%%%%%%%%%%%%%%%%%%%%%%%%%

%%%%%%%%%%%%%%%%% BODY OF PAPER %%%%%%%%%%%%%%%%%%

\section{Introduction}
\label{sec:intro}

There were 10 \ac{BBH} detections \citep{abbott2016binary,LVC2018O2} and a \ac{BNS} \citep{abbott2017gw170817} in the first and second observing runs of the \ac{aLIGO} and Virgo gravitational-wave detectors. The intrinsic rate of \ac{BBH} mergers is currently estimated by the LIGO-Virgo collaboration at $24$--$112$\,$\rm Gpc^{-3} yr^{-1}$ , whereas for \acp{BNS} it is $110$--$3840$\,$\rm Gpc^{-3} yr^{-1}$ \citep{2018arXiv181112940T}. These intrinsic rate estimates depend on the assumed shape of the mass and rate distribution of the \ac{DCO} mergers, which remains uncertain. Multiple possible stellar origins exist for \acp{DCO} such as dynamical capture in open/globular/nuclear clusters, Lidov-Kozai resonances in hierarchical triples, chemically homogeneous evolution in compact stellar binaries, and mergers of primordial black holes \citep[see][for reviews]{2016GReGr..48...95M,2018arXiv180605820M,giacobbo2018progenitors}. We focus on the merger rate of \acp{DCO} that come from isolated binary evolution. It appears that most of the massive stars (M\,$> 8$ M$_\odot$) in the field are born in binaries \citep{2012KiminkiCygnusOB2, 2013SanaVLTFOstar, 2017MoePQ}. Once formed, these isolated binaries evolve without external influences and a fraction becomes \acp{DCO}. However, the exact physics of stellar and binary evolution and the resulting rates of \ac{DCO} mergers are still uncertain \citep[e.g.,][]{2015DominikIIIGWrates, 2016MNRAS.462.3302E, 2018MNRAS.481.1908K, chruslinska2018influence}.

The evolution of massive stars takes a few million years, but their inspiral as \acp{DCO} can span years to billions of years \citep[e.g.,][]{1998ZwartDNS,2002BelczynskiBBH, 2016MNRAS.462.3302E, 2017MapelliCosmicBH}. The detected mergers could therefore have formed at very high redshifts. Observations show that the \ac{SFR} changes significantly as a function of redshift \citep{madau2014cosmic}. At redshifts $z \gtrsim 2$ the \ac{SFR} estimates become increasingly more sensitive to the assumed extinction, which is uncertain \citep{madau2014cosmic, strolger2004hubble}. The \ac{SFR} determines the amount of stellar binaries formed and hence introduces an uncertainty on the rate of \ac{DCO} formation.

Metallicity, and particularly the fraction of iron in the star at birth, significantly impacts the rate of mass loss through line-driven winds.  Consequently, it  has a significant effect on the \ac{DCO} mass distribution and merger rate \citep{belczynski2010maximum, stevenson2017formation, 2018MNRAS.474.2959G}.  The metallicity of star forming gas depends on redshift, as subsequent generations of stars enrich the inter-stellar medium through winds and explosions with metals formed during their evolution.  Galaxy catalogs, such as the Sloan Digital Sky Survey \citep{tremonti2004origin}, show that there is an empirical correlation between the galaxy stellar mass and the mean metallicity of the galaxy. Furthermore, the \ac{GSMF} and the \ac{MZ}-relation evolve with redshift.   Different calibrations or galaxy samples lead to different results \citep{savaglio2005gemini,2008KewleyMZ, furlong2015evolution}. Thus there is not only uncertainty in the overall \ac{SFR} but also in the distribution
of the metallicities in the star forming gas. Combined, these result in an uncertainty in the \ac{MSSFR},
which affects estimates of the rates and properties of \ac{DCO} mergers.

\subsubsection*{Aim and overall method}
Our aim is to assess how the uncertainty in the \ac{MSSFR} affects 
predictions for the rate and distributions of \ac{DCO} mergers.  In this section, we introduce the key steps in the calculation of the redshift-dependent \ac{DCO} merger distribution and the rate of detectable \ac{DCO} mergers.

The time it takes for a binary to evolve its stars and then merge at $t_\mathrm{m}$ as a \ac{DCO} due to the emission of gravitational waves is called the delay time ($ t_\mathrm{delay}$).   The formation time $t_\mathrm{f}$ is related to the merger time $t_\mathrm{m}$ by $t_\mathrm{f} = t_\mathrm{m} - t_\mathrm{delay}$.  We can calculate the rate of mergers at any given time as
\begin{equation}
\begin{split}
\label{eq:intrinsic}
{} & \dfrac{d^3 N_\mathrm{merge}}{d t_\mathrm{s}d V_\mathrm{c} dM_\mathrm{chirp}} (t_\mathrm{m}) = \int dZ \int dt_\mathrm{delay} \\
& \dfrac{d^3 N_\mathrm{form}}{d t_\mathrm{delay}d M_\mathrm{SFR} d M_\mathrm{chirp}} (Z) \dfrac{d^3 M_\mathrm{SFR}}{d t_s\ d V_c\ d Z}(t_\mathrm{f} = t_\mathrm{m} - t_\mathrm{delay}) , 
\end{split}
\end{equation}
where $Z$ is the metallicity, $z$ is the redshift, $t_s$ is the time in the source frame of the merger, and $V_\mathrm{c}$ is the comoving volume. The first term in the integrand is the number of \acp{DCO} per unit star forming mass $M_\mathrm{SFR}$ per unit delay time and per unit chirp mass $M_\mathrm{chirp} = M_1^{3/5} M_2^{3/5} (M_1+M_2)^{-1/5}$, where $M_1, M_2$ are the individual compact object masses. We compute this first term over a grid of metallicities by running the COMPAS population synthesis code. The second term is the \ac{MSSFR} at the birth of the binary per unit time, volume, and metallicity, which we model analytically. 

The second step is to calculate the distribution of observable \ac{DCO} mergers. We do this by converting $t_{\mathrm{m}}$ to a redshift $z$ and integrating the entire visible volume in shells of thickness $dz$. At each redshift we calculate the probability of detecting a binary ($P_\mathrm{det}$) given its chirp-mass ($M_\mathrm{chirp}$) and luminosity distance ($D_\mathrm{L} (z)$). The total observable merger distribution is then
\begin{equation}
\begin{split}
\label{eq:detections} 
 	\dfrac{d^2 N}{d t_\mathrm{obs} d M_\mathrm{chirp}}  = {} &\int_{0}^{z_\mathrm{max}} d z \dfrac{dt_s}{d t_\mathrm{obs}} \dfrac{dV_\mathrm{c}}{dz} \times \\
  	& \dfrac{d^3 N_\mathrm{merge}}{dt_{\mathrm{s}}\ dV_\mathrm{c} dM_\mathrm{chirp}}(z)
P_\mathrm{det}(M_\mathrm{chirp}, D_\mathrm{L})\ , 
\end{split}
\end{equation}
where $d V_c/dz$ is the differential comoving volume as a function of redshift and $dt_\mathrm{s} / dt_\mathrm{obs} = 1/(1+z)$ translates the rate to the observer frame \citep[e.g.][]{hogg1999distance}. We assume a flat cosmology with $\Omega_\mathrm{M}= 0.308 $ and a Hubble constant of $H_0=67.8$ km s${-1}$ Mpc$^{-1}$ \citep{ade2016planck}. Altogether this general method is similar to
works such as \citet{langer2006collapsar}, \citet{2013DominikIIRates}, \citet{mandel2016merging},
\citet{2016MNRAS.462.3302E}, \citet{2017ApJ...840...39M} and \citet{chruslinska2018influence}.

We sequentially solve the aforementioned equations, structuring the paper as follows:
\begin{itemize}
 \item[--]{ Section~\ref{popsynth}: COMPAS population synthesis code  \\
 We create a large sample of \acp{DCO} from a broad range of metallicities using the rapid population synthesis element of the COMPAS suite.
 We briefly describe the model assumptions
 used to evolve our massive stellar binaries.}
 \item[--]{ Section~\ref{COMPAS}: $ \frac{d^3N_\mathrm{form}}{d t_\mathrm{delay}d M_\mathrm{SFR} d M_\mathrm{chirp}}$ - \ac{DCO} population  \\
 We show the results of our population synthesis of \acp{DCO}. 
 We describe some of the key features such as their mass distribution at different formation metallicities in our simulation. We describe the three main formation channels for \acp{BBH}. 
 We find a significant number of \acp{BBH} merging without experiencing a common-envelope event.}
 \item[--]{ Section~\ref{MSSFR}: $\frac{d^3 M_\mathrm{SFR}}{d t_{s} d V_{c} d Z}$ - MSSFR   \\
 We combine observations and simulations of galaxy stellar mass distributions with 
mass -- metallicity relations to construct a \ac{MSSFR}. These different prescriptions introduce an uncertainty into our \ac{DCO} merger rate distributions.
We propose a parametrised, smooth  metallicity distribution, which facilitates the exploration of
 the \ac{MSSFR} parameter-space
 }
 \item[--]{ Section~\ref{CosmicDCOrate}: $\frac{d^3N_\mathrm{merge}}{dt_s dV_c dM_\mathrm{chirp}}$ - \ac{DCO} Merger Distributions  \\
 We calculate the redshift-dependent \ac{DCO} distribution by convolving the \ac{MSSFR}
 with our \ac{DCO} population. We find that variation in \ac{MSSFR} prescriptions
 significantly affects both the total rate and mass distributions of \acp{DCO} mergers.}
\item[--]{ Section~\ref{GWdetections}: $\frac{d^2 N}{d t_\mathrm{obs} d M_\mathrm{chirp}}$ - Gravitational-Wave Detections \\
We apply selection effects of gravitational-wave detectors to our cosmic \ac{DCO} populations. 
From this we get both rate and mass distributions of detectable \ac{BBH} mergers for different \ac{MSSFR} prescriptions. We use a Bayesian approach to compare the predictions of different \ac{MSSFR} models against the observed sample of gravitational waves from \ac{BBH} mergers. We find that the \ac{MSSFR} significantly affects
the predicted rate of gravitational-wave events from \ac{BBH} mergers.}
\item[--]{ Section~\ref{discussion}: Discussion and conclusion  \\
We review our findings and discuss future prospects.}

\end{itemize}

%%%%%%%%%%%%%%%%%%%%%%%%%%%%%%%%%%%%%%%%%%%%%%%%%%%%

\section{COMPAS Population Synthesis Code}
\label{popsynth}

We generate our population of \acp{DCO} by modelling isolated binary evolution with the population synthesis code COMPAS \citep{stevenson2017formation, 2018JimInference, 2018AVGDNS, 2019arXiv190402821S}.  We use Monte Carlo simulations to empirically estimate the rate density of \acp{DCO} per unit star forming mass in delay time and chirp mass at each simulation metallicity:
$$\dfrac{d^3 N_\mathrm{form}}{d t_\mathrm{delay}d M_\mathrm{SFR} d M_\mathrm{chirp}} (Z, t_\mathrm{delay}, M_\mathrm{chirp}).$$
In this section we briefly describe the parameter space of our simulation and our model assumptions for isolated binary evolution.
The data will be made publicly available at \url{http://compas.science}.

\subsection{Initial Distributions}
\label{subsec:initial_distributions}

The five initial conditions that describe a stellar binary are: the primary $m_1$ and secondary $m_2$ masses, the orbital separation $a$, the orbital eccentricity $e$, and the metallicity of the stars $Z$ at \ac{ZAMS}. The mass of the initially more massive star, the primary, is drawn from an \ac{IMF} according to \citet{2001MNRAS.322..231K}. The mass of the initially less massive secondary star is given by 
\begin{equation} 
\label{eq:mass_ratio}
	m_{2} = m_{1} \times q, 
\end{equation} 
where $q$ is the initial mass ratio ($0 < q < 1$). We draw the mass ratio $q$ from a flat distribution \citep{2012SanaInteraction}. We assume that the distribution of separations is flat-in-the-log ($0.1 < a/\mathrm{AU} < 1000$) \citep{opik1924statistical} and the orbits are all  circular at birth. We assume that these distributions are both independent of each other as well as independent of metallicity. Recent studies such as \citet{2017MoePQ} suggest that the initial distributions might be correlated. \cite{2015MinkInitial,2018KlenckiInitial} found that varying initial condition distributions affects \ac{DCO} merger rates by factors of $\lesssim 2$.

For the metallicities of the binaries we use 30 grid points spread uniformly in log-space over a broad range  of metal mass fractions $\rm 0.0001 \leq Z \leq 0.03$. We evolve three million binaries with a total star forming mass of the order of $\rm 6.5 \times 10^7 \, M_{\odot}$ per grid-point. 

To optimise the number of compact objects per binary simulated, whilst still leaving enough room in the parameter space to avoid boundary effects, we draw primaries with masses equal or bigger than 5~$\mathrm{M}_{\odot}$ (this represents a very naive version of importance sampling introduced by \citealt{2019arXiv190500910B}). Our upper mass limit is 150 solar masses. In this mass range the power index of the \ac{IMF} equals -2.3. Hence we need to correct for the `true' amount of mass evolved in all stars (both single and binary). We calculate this by assuming a binary fraction of 70 per cent and a flat mass ratio for all stellar masses \citep{2012SanaInteraction}. This results in a total star forming mass per metallicity grid point of $\sim \rm 3.1\times10^8\,  M_{\odot}$. It is this star forming mass that we use as our normalisation $d M_\mathrm{SFR}$.

\subsection{Single Stellar Models}
\label{subsec:single_stellar_models}

Stellar evolution in COMPAS is based on the stellar models by \citet{pols1998SSE}. We use analytical fits to these models  by \citet{ hurley2000comprehensive, hurley2002evolution} to rapidly evolve binaries. Our wind mass loss rates for stars with temperatures below 12500 K  are prescribed by \citet{hurley2000comprehensive} and references therein. For hot massive stars ($\rm T > 12500K$) we use the wind mass loss rates by \citet{2001A&A...369..574V} as implemented in \citet{belczynski2010maximum}. There is a region in the Hertzsprung-Russell diagram at low effective temperatures and high luminosities in which no stars are observed. The boundary of this region is called the Humphreys-Davidson limit \citep{1994HumphreysLBV}. If a star enters this region we apply an additional wind mass loss rate of $\rm 1.5 \times 10^{-4}\ M_{\odot}\ yr^{-1}$ \citep{belczynski2010maximum}. From here onwards we refer to these winds as \ac{LBV} winds. 

\subsection{Mass Transfer Stability}
\label{stability}

The Roche lobe of a star defines the volume within which the self gravity of the star exceeds the tidal pull of its companion.  We use the approximation of \citet{eggleton1983approximations} for the Roche lobe radius. When a star expands, its radius may exceed its Roche lobe.  At this moment, the star commences mass transfer onto the companion, \ac{RLOF}. If mass transfer results in the star further exceeding its Roche-lobe then the \ac{RLOF} is unstable. We evaluate dynamical instability by comparing the radial response of the Roche-lobe  to mass transfer  $d\log(R_{{L}})/d\log(m)$ against the response of the stellar radius to mass transfer $  d\log(R_{{*}}) /d\log(m)$ \citep{1972AcA....22...73P,1987HjellmingRadii, 1997SobermanRadii}. We approximate the radial response of the star depending on its stellar type. The stellar types are defined in \citet{hurley2000comprehensive}.

\begin{itemize}
 \item[--]{\ac{MS}:\\ We use $d\log(R_*)/d\log(m)=2.0$ for core hydrogen burning stars.}
\item[--]{\ac{HG}:\\
We use $d\log(R_*)/d\log(m)=6.5$ for so-called \ac{HG} stars.  Both \ac{MS} and \ac{HG} approximations follow our models in \citet{2018AVGDNS}, based on the work by \citet{2015GeRadii}.  More detailed models based on the evolutionary phase of the star and the amount of mass loss have been explored by \citet{2015GeRadii, 2011ApJ...739L..48W, 2017MNRAS.465.2092P}.
}
\item[--]{Convective stars:\\
We use fits from \citet{1987HjellmingRadii, 1997SobermanRadii} for the radial response to adiabatic mass loss of all evolved stars beyond \ac{HG}.   These fits are based on condensed polytropes for deeply convective stars and depend on the mass fraction of the core compared to the total mass of the star \citep{1987HjellmingRadii}.  We will investigate the applicability of these approximations in future work \citep{NeijsselPrep}. }
\item[--]{Stripped stars:\\
We make a special exception for mass transfer from exposed helium cores.  We define this mass transfer to always be dynamically stable, yielding ultra-stripped stars based on \citet{2015TaurisDNS, 2017TaurisDNS}. \citet{2018AVGDNS} found that this assumption is necessary in order to recreate the observed Galactic double neutron stars in our models.}
\end{itemize}

\subsubsection{Stable Mass Transfer}
\label{stable}

If the mass transfer is dynamically stable, the companion star accretes a fraction $\beta$ of the mass lost by the donor.  In our model, this mass transfer efficiency $\beta$ depends on the ratio of the thermal timescales $t_\mathrm{th}$ of the stars $\beta = \mathrm{min} \left(1, C \times t_\mathrm{th_1}/t_\mathrm{th_2} \right)$, where $0 \leq \beta \leq 1$, and $C = 10$ to allow for accretor radial expansion while adjusting to mass transfer \citep{1972AcA....22...73P, hurley2002evolution, 2015SchneiderMT}. Any mass that is not accreted leaves the system instantaneously, taking away the specific angular momentum of the accretor \citep{hurley2002evolution}. For degenerate objects we assume the accretion is Eddington-limited, which results in a highly non-conservative mass transfer phase with $\beta \approx 0$.

\subsubsection{Unstable Mass Transfer}
\label{subsubsec:unstable_mass_transfer}

If the mass transfer is unstable the envelope of the donor enfolds the entire  binary in a common-envelope event \citep{1976PaczynskiCEE}. This is a complex phase and we parametrise it in the so-called `$\alpha$--$\lambda$' formalism (see \citet{ivanova2013common} for a review). During a common-envelope event the two stars spiral in due to friction with the envelope and lose orbital energy and angular momentum. This loss of orbital energy can heat up and expel the envelope. To see if a binary is able to expel the common envelope, we compare the orbital energy against the binding energy of the envelope of the star \citep{webbink1984double}. The efficiency $\alpha$ of converting orbital energy into heating up the envelope can vary \citep{1988LivioAlpha}. We assume that all of the orbital energy goes into expelling the envelope (i.e. $\alpha=1$). The binding energy of the envelope depends on the stellar structure of the star and is parametrised by $\lambda$ \citep{1990KoolLambda}. Our choices
of $\lambda$ are based on the binding energy fits by \citet{2010XuLambda} as implemented by \citet{2012DominikCEE}.

Within the common envelope we define two scenarios for donor stars which are on the Hertzsprung-gap, following \citet{belczynski2007rarity}. In the `optimistic` scenario we evaluate the common-envelope evolution for Hertzsprung-gap stars using the `$\alpha$--$\lambda$' prescription. In the `pessimistic` scenario we assume that unstable mass transfer from Hertzsprung-gap donors always results in a merger. The latter will therefore decrease the number of \acp{DCO} compared to the optimistic assumption.  Common-envelope events with \ac{MS} donors are assumed to lead to a prompt merger in all variations.

\subsubsection{Supernovae}
\label{subsubsec:supernovae}

We use the 'delayed' model of \citet{2012FryerRemnants} to determine the remnant mass from the pre-\ac{SN} mass of the star and its carbon-oxygen core.  This model avoids an enforced mass gap between \acp{NS} and \acp{BH} (see also evidence that a mass gap is not consistent with microlensing observation unless \acp{BH} are assumed to receive substantial natal kicks \citep{WyrzywkoskiMandel:2019}). The explosion can be asymmetric and as a result impart a kick on the formed remnant. The kicks are drawn from a Maxwellian distribution with a one-dimensional standard deviation $\rm \sigma = 265\ km~s^{-1}$ based on the observations of isolated pulsars \citep{2005HobbsKicks}. If the progenitor either experiences an electron capture supernova or is ultra stripped by a \ac{NS} companion, we lower the one-dimensional kick parameter to 30 $\rm km\ s^{-1}$ \citep{2002PfahlECSN, 2004PodsiadlowskiKick,2015TaurisDNS, 2017TaurisDNS,2018AVGDNS}.  The fraction $\rm f_{b}$ of mass that falls back onto the newly born compact object is prescribed by \citet{2012FryerRemnants}. All of the ejecta falls back ($\rm f_{b}$=1) for carbon-oxygen core masses above 11~$\rm M_\odot$. This natal kick is proportionally reduced based on the fallback fraction according to 
\begin{equation}
\label{eq:kick}
	V_\mathrm{kick} = (1-f_{b}) V_\mathrm{kick,drawn}.
\end{equation}
%

%\clearpage

%%%%%%%%%%%%%%%%%%%%%%%%%%%%%%%%%%%%%%%%%%%%%%%

\section{\ac{DCO} Population}
\label{COMPAS}

In this section we describe the three main \ac{BBH} formation channels.  We focus on \acp{BBH} because they are the most common \acp{DCO} among already observed gravitational-wave events. More information on \acp{BNS} can be found in \citet{2018AVGDNS} and the channels for \acp{BHNS} are left for another study. We aslo show the metallicity, mass, mass-ratio, and delay time distributions for our model \ac{DCO} population.

A 30\ $\rm M_\odot$\ + 30\ $\rm M_\odot$ circular \ac{BBH} needs a separation of $\lesssim 45\  R_\odot$ to merge in the age of the Universe, whereas the progenitor stars can expand up to hundreds of solar radii \citep{2018arXiv180605820M}. Therefore, progenitors of \acp{DCO} are expected to interact.  This is not unlikely to happen for massive stars in binaries: observations show that most
massive stars are likely to interact with a companion \citep{2012KiminkiCygnusOB2,2012SanaInteraction}. Only a small fraction of interacting massive binaries will form merging \acp{DCO}.  This requires stars to avoid merger during mass transfer; to have sufficient mass to form compact objects; the binary must remain bound through \acp{SN}; and after the formation of a \ac{DCO}, the binary must be tight enough to merge within the age of the Universe and create a detectable gravitational-wave event. Our main goal is to evaluate the \acp{DCO} that we can detect as gravitational-wave sources; hence we are only interested in the systems that merge within the age of the Universe. The results shown below assume the pessimistic common-envelope assumption.  The optimistic assumption currently over-predicts the rates of \acp{BBH} \citep[e.g.][]{2012DominikCEE,belczynski2016first}; we show results using the optimistic assumption in Appendix~\ref{Bayes}.

\subsection{BBH Formation Channels}
\label{subsec:channels}

In our simulations, 97\% of all the \acp{BBH} form through one of three distinct channels. 
Here we briefly summarise the evolutionary phases of the three main formation channels and the percentages of systems that remain
after having experienced a given even/phase. 
The exact percentage and the ratio of the formation channels depends slightly on metallicity (see also figure~\ref{fig:yield}).
We focus on the systems that evolved at a metallicity of $Z=0.1 Z_\odot$,\footnote{In this study, we define the solar metallicity mass fraction as $Z_{\odot}=0.0142$ and the solar oxygen abundance as $ \log_{10}[\mathrm{O/H}]_{\odot} + 12 = 8.69$ based on \citet{2009AsplundMetallicity}; see appendix \ref{subsec:solar_values} for details.}
and the percentage refers to the number of systems
remaining divided by all of the systems evolved at this metallicity.
 \\
 \\
-- \textit{Channel I} -- \\ 
This is the dominant, `classical' channel of \ac{BBH} formation as described in, e.g., \citet{1973A&A....25..387V, TutukovYungelson:1993,Lipunov:1997,2002ApJ...571L.147B, belczynski2016first, stevenson2017formation}.\\
\textit{-Stable mass transfer-} The primary star expands sufficiently to engage in an episode of mass transfer (51.71\%). The majority of these first mass transfer episodes will happen between a post-\ac{MS} primary star and a \ac{MS} companion (49.26\%). The mass transfer is stable and strips
the hydrogen envelope from the primary, leaving an exposed helium core with a main-sequence companion star (23.06\%). \\
\textit{-First supernova-} The exposed core is both massive enough to collapse into a \ac{BH} and the binary survives the supernova (2.66\%).\\
\textit{-Unstable mass transfer-}
The secondary star evolves and starts an episode of dynamically unstable mass transfer resulting in a common envelope (0.87\%). The system is able to expel the envelope leaving a tighter binary (0.50\%). \\
\textit{-Second supernova-}
The secondary also collapses into a \ac{BH} and the binary system survives the second supernova (0.43\%).\\
\textit{-DCO merger-} The resulting \ac{BBH} is then able to merge within the age of the Universe due to the emission of gravitational waves, which leaves 0.24\% of all our evolved binaries merging as \acp{BBH}. Allowing for the optimistic common-envelope assumption, in which \ac{HG} donors can survive a dynamically unstable mass transfer episode, increases the number of \acp{BBH} in this channel  (0.39\%).
\\
\\
-- \textit{Channel II} -- \\
The second channel is similar to the `classical' channel and goes through the same steps until the episode of mass transfer initiated by the secondary, which is dynamically stable in channel II.
\textit{-Stable mass transfer-}  see channel I. \\
\textit{-First supernova-} see channel I.\\
\textit{-Stable mass transfer-}
The secondary starts mass transfer as a post-MS star. The mass transfer is now dynamically stable and \textit{does not result in a common-envelope phase} (1.35\%). \\
\textit{-Second supernova-} The secondary collapses into a \ac{BH} without disrupting the binary (1.02\%). \\
\textit{-DCO merger-}
Even without the common-envelope phase the \ac{BBH} hardens (reduces orbital separation)  sufficiently during the second mass transfer episode to spiral in and merge within the age of the Universe (0.15\%).

Compared to \citet{stevenson2017formation} we changed the radial response of \ac{HG} donors to mass loss (see Sec.~\ref{stability}). In combination with our prescription for the angular momentum lost during non-conservative mass transfer onto a compact-object primary (see Sec.~\ref{stable}), the mass transfer is on average now stable for mass ratios up to $m_{\rm donor}/m_{\rm accretor}=4.5$.
Mass transfer from such donors that are significantly more massive than accretors can substantially harden the binary \citep{2017HeuvelStability}. With the increased stability these mass ratios are sufficiently extreme to allow the \ac{BBH} to merge within the age of the Universe. 

The stability of the second episode of mass transfer acts as a bifurcation point between channel I and channel II. Currently, channel II only happens for \ac{HG} donors in our models, since we treat core-helium-burning donors as fully convective, making mass transfer from them less dynamically stable. The potential importance of this formation channel and the stability of mass transfer is discussed in previous studies  \citep[see for example][and references therein] {2017PavlovskiiStability,2017HeuvelStability}. 
\\
\\
-- \textit{Channel III} --\\
The third channel for forming \acp{BBH} is similar to the double-core common-envelope channel introduced by \citep{1995ApJ...440..270B,Dewi:2006bx}.
\textit{-Unstable mass transfer-} . In this scenario both stars evolved beyond the \ac{HG} before engaging in an episode of mass transfer (1.40\%). This mass transfer is dynamically unstable (1.27\%) and the binary survives the  common-envelope ejection (0.71\%). However, unlike the similar  formation channel for \acp{BNS} \citep{2018AVGDNS}, there is no further episode of mass transfer.\\
\textit{-Two supernovae-} Both stars collapse in supernovae (non-simultaneously); 0.04\% of binaries remain bound as a \ac{BBH}.\\
\textit{-DCO merger-} The \ac{DCO} spirals in due to the emission of gravitational waves. In the end 0.03\% of all binaries evolved go through this channel and merge within the age of the Universe. 
\\
\\

The remaining three per cent of \acp{BBH} form through alternative channels. These include systems which have an additional moment of mass transfer after a common-envelope phase, or systems where the first moment of mass transfer is started by the secondary after the primary's supernova kick fortuitously tightened the binary.

%\clearpage

%%%%%%%%%%%%%%%%%%%%%%%%%%%%%%%%%%%%%%%%%%%%

\subsection{Yield Per Metallicity}
\label{subsec:yield}

\begin{figure}
\includegraphics[width=\columnwidth]{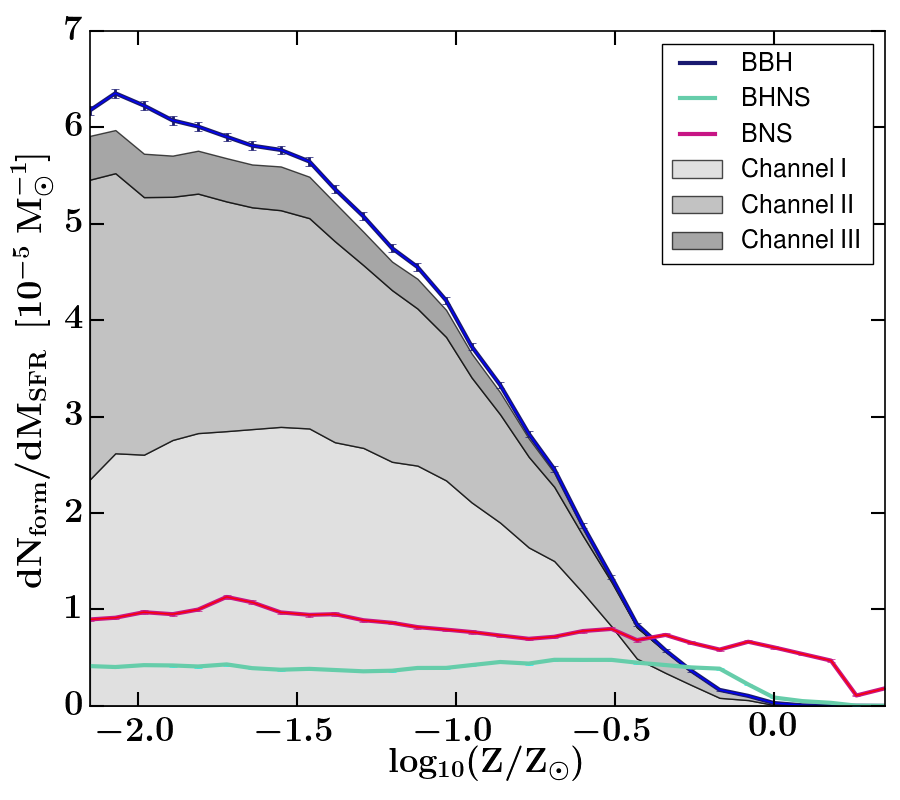}
   \caption{Yield of double compact objects with $t_\textrm{delay}<14$ Gyr per unit star forming mass from COMPAS population synthesis. \acp{BBH} in blue, \acp{BHNS} in mint and \acp{BNS} in red. The curve under the \ac{BBH} yield is shaded by the contribution of each channel (the white residual is due to rare alternative channels). The error bars show the sampling uncertainty of each simulation.}
    \label{fig:yield}
\end{figure}

The yield of merging \acp{DCO} per unit star forming mass depends on the star formation metallicity, as shown in figure \ref{fig:yield}.  As previously pointed out by \citet{belczynski2010maximum,2018MNRAS.474.2959G,2018SperaSEVN}, \ac{BBH} yield is particularly sensitive to metallicity with a steep decline in \ac{BBH} production at higher metallicities.  Therefore, while \acp{BBH} are the dominant form of merging \acp{DCO} at sub-solar metallicities, they are more rare than \acp{BNS} and \acp{BHNS} at super-solar metallicities.

At higher metallicities, higher wind mass loss rates prevent the growth of the carbon-oxygen core
 \citep{belczynski2010maximum,2015MNRAS.451.4086S,stevenson2017formation}, leaving a less massive remnant.  This affects the natal kicks imparted on the \acp{BH}. In the prescription of \citet{2012FryerRemnants}, stars with lower carbon-oxygen cores eject a larger fraction of their mass which results in larger natal kicks (see Eq.~\ref{eq:kick}).

Therefore, we expect more potential \ac{BBH} progenitors to be disrupted at higher metallicities. A smaller simulation without natal kicks does show a shallower drop-off of the \ac{BBH} yield at higher metallicities. Nonetheless, there is still a lower yield at higher metallicities.  This is largely due to the widening of binaries at higher metallicity, both directly through wind-driven mass loss and indirectly because reduced envelope masses limit the amount of orbital hardening during common-envelope ejection or stable mass transfer.  In fact, if \ac{BH} natal kicks are set to zero and all \acp{BBH} are accounted for, not just those merging in the age of the Universe, the \ac{BBH} yield becomes almost independent of metallicity.

Lower-mass \ac{NS} progenitors have lower mass loss rates, so the envelope mass is less sensitive to metallicity; moreover, their natal kicks are generally uncorrelated with metallicity. Hence it is not surprising that the yield of \acp{BNS} per unit solar mass evolved is less sensitive to metallicity, as also found by \citet{Giacobbo:2018hze}.

%%%%%%%%%%%%%%%%%%%%%%%%%%%%%%%%%%%%%%%%

\subsection{Total Mass Distribution}
\label{subsec:total_mass_distribution}

\begin{figure}
\includegraphics[width=\columnwidth]{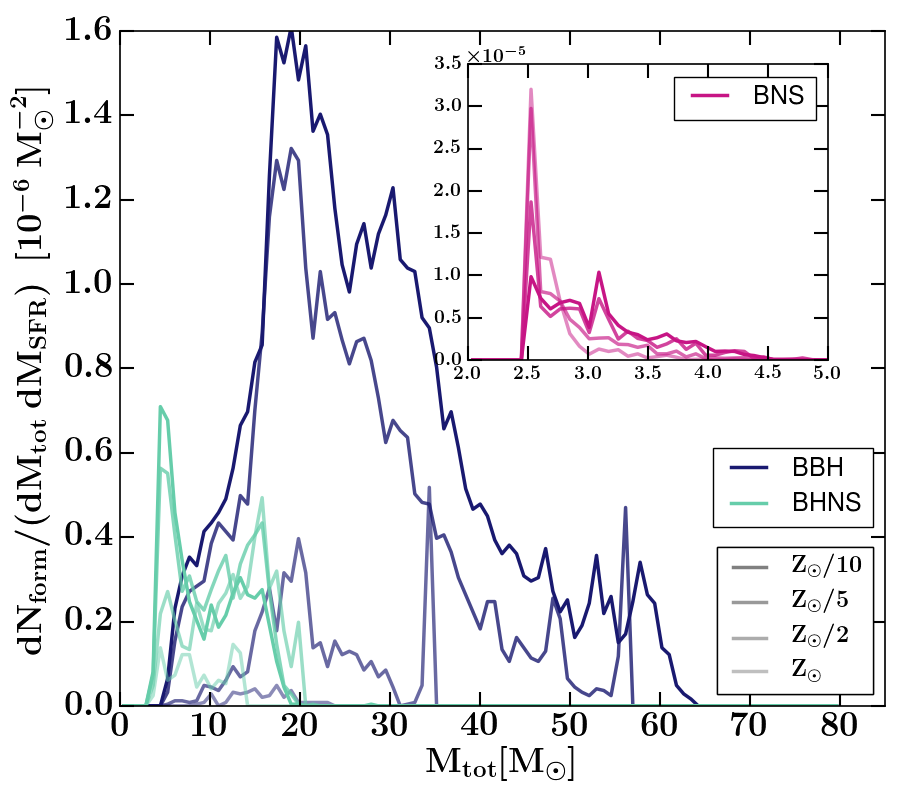}
   \caption{Total mass distributions for \acp{BBH} in blue, \acp{BHNS} in mint and \acp{BNS} in red from
   COMPAS simulations for a tenth, a fifth, a half, and solar metallicity (dark to light shade), for \acp{DCO} merging in $t_\textrm{delay}<14$ Gyr. 
   The integral under the curve is the yield plotted in Fig.~\ref{fig:yield}.
   Higher metallicities yield lower total \ac{DCO} masses, particularly for \acp{BBH}.}
    \label{fig:Mtot}
\end{figure}

Figure~\ref{fig:Mtot} shows the total mass distributions of \acp{DCO} merging within the age of the Universe for several metallicities.   As discussed in the previous section, lower-metallicity stars with reduced wind-driven mass loss rates leave more massive remnants.  For all metallicities the bulk of the \ac{BBH} total masses lie between 15 \& 35~$\rm M_{\odot}$. More massive \acp{BBH} are suppressed by the \ac{IMF} and wind-driven mass loss.  The most massive binary black hole formed at a given metallicity is a function of both our assumptions about wind mass loss in massive stars, and our remnant prescription \citep[these simulations do not include (pulsational) \ac{PISN} -- see][]{2019arXiv190402821S}. Meanwhile, \acp{BH} with low masses get large kicks in the \citet{2012FryerRemnants} prescription, and are therefore less likely to remain bound and form a \ac{BBH}, explaining a dearth of \acp{BBH} with total mass below $15\,\mathrm{M}_\odot$. The `delayed' \citet{2012FryerRemnants} remnant prescription does not enforce a mass gap between \acp{NS} and \acp{BH}, so we find some \acp{BBH} with total masses below $10\,\mathrm{M}_\odot$ in our simulations, although these are relatively rare.

The presence of spikes in \ac{BBH} masses, particularly in the highest mass bin at $Z=0.5 Z_\odot$ and $Z=0.2 Z_\odot$, are due to mass loss prescriptions, particularly \ac{LBV} winds, that map a range of \ac{ZAMS} masses to a single remnant mass (see Appendix.~\ref{AppRemnant}). Similar features have been found in \citet{2015DominikIIIGWrates}.

For \acp{BNS} we recover a similar total mass distribution as in \citet{2018AVGDNS}.  As discussed in \citet{2018AVGDNS}, this distribution, driven by the \citet{2012FryerRemnants} prescription, does not match the observed distribution of Galactic \acp{BNS}.  For example, in our model, \acp{BNS} have total masses in the range $2.5$--$5.0$\,M$_\odot$, while observed Galactic \acp{BNS} with precise mass measurements have total masses in the narrower range 2.5--3.0\,M$_\odot$ \citep[][]{Farrow:2019xnc}.  

%%%%%%%%%%%%%%%%%%%%%%%%%%%%%%%%%%%%%%%%%%%%%%%%%%%%

\subsection{Delay Times}
\label{subsec:delay_times}

\begin{figure}
\includegraphics[width=\columnwidth]{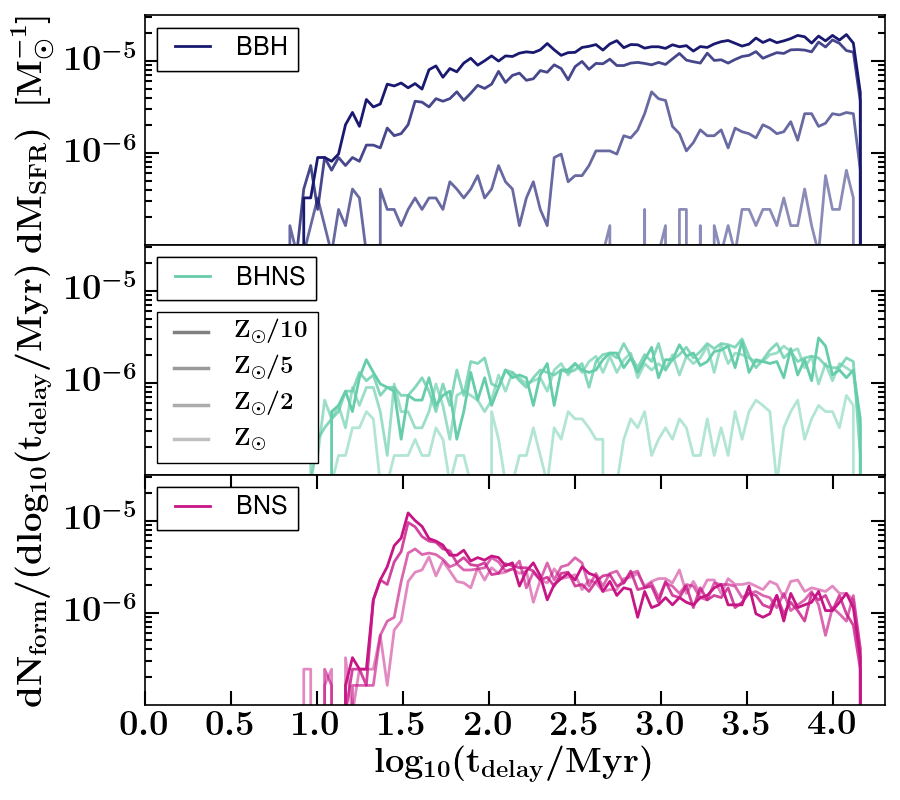}
   \caption{Delay time distributions up to $t_\textrm{delay}=14$ Gyr for \acp{BBH} in blue, \acp{BHNS} in mint and \acp{BNS} in red from
   COMPAS simulations for a tenth, a fifth, a half, and solar metallicity (dark to light shade).
   }
   \label{fig:tDelay}
\end{figure}

The delay time is the time from the formation of the stars to their merger as a \ac{DCO}. 
We follow \citet{peters1964gravitational} to estimate the time from \ac{DCO} formation to merger through the emission of gravitational waves. The most massive binaries with the smallest separation at the formation of the \ac{DCO} have the shortest inspiral times.  The delay time distribution is roughly flat-in-the-log for all \acp{DCO} (see Fig.~\ref{fig:tDelay}). Furthermore, for the pessimistic assumption it is not very sensitive to metallicity.  These findings are similar to \citet{2012DominikCEE} and \citet{2017MapelliCosmicBH}.  

%%%%%%%%%%%%%%%%%%%%%%%%%%%%%%%%%%%%%%%%%%%%%%%%%%%%

\subsection{Mass Ratios}
\label{subsec:mass_ratios}

\begin{figure}
\includegraphics[width=\columnwidth]{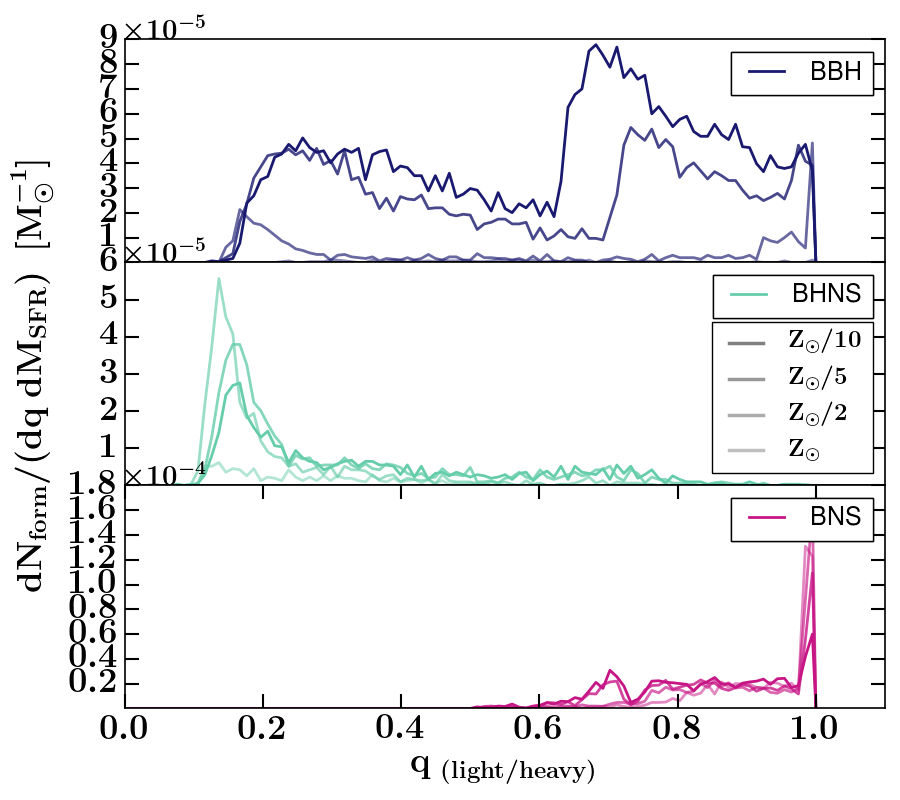}
   \caption{Mass ratio distributions for \acp{BBH} (blue), \acp{BHNS} (mint) and \acp{BNS} (red) merging in $t_\textrm{delay}<14$ Gyr from
   COMPAS simulations for a tenth, a fifth, a half, and solar metallicity (dark to light shade).
   For \acp{BBH} channel II contributes more at lower metallicities and therefore the mass ratio distribution at $0.1 Z_\odot$
   has a prominent feature around $q \approx 0.65$.}
    \label{fig:massRatio}
\end{figure}

Figure \ref{fig:massRatio} shows the mass ratio distributions of \acp{DCO} merging within the age of the Universe at several metallicities. It is clear that the distributions differ between different types of \acp{DCO} and depend on metallicity.

The mass ratios of \acp{BNS} exhibit two dominant peaks. The mass ratio is close to unity if both \acp{NS} have the lowest allowed remnant mass from iron-core-collapse supernovae in the \citet{2012FryerRemnants} `delayed' prescription.  The other peak is the ratio between this lowest iron-core-collapse remnant mass and the fixed remnant mass from electron-capture supernovae, 1.26~$\rm M_{\odot}$ in our model.  The prevalence of these peaks is enhanced by binary interactions.  We do not expect extreme mass ratios given the limited spread in possible \ac{NS} masses.

The \acp{BHNS} favour more extreme mass ratios. The average \ac{NS} mass is 1.2~$\rm M_{\odot}$ and the threshold between \ac{NS} and {BH} is 2.5$\rm M_{\odot}$ in our models. This already results in a mass ratio of 0.5, but most of the \acp{BH} are heavier. Further details are outside the scope of this study.

The mass ratio distribution of \acp{BBH} depends on the formation channel.  The classical channel I with a common-envelope phase occurs for a broad range of mass ratios between the donor star and the accreting \ac{BH}. This channel yields a relatively flat mass-ratio distribution.  Meanwhile, channel II, in which the mass transfer onto the \ac{BH} is dynamically stable, has an upper limit of 4.5 for the mass ratio between the donor and the \ac{BH} accretor.   Mass ratios close to this limit are preferred as they provide the most orbital hardening.  After this mass transfer, the stripped donor star collapses into a \ac{BH}.  This results in a \ac{BBH} mass ratio around $q \approx 0.6$.  If such an additional peak is observed in the mass ratio distribution of gravitational-wave events, its prominence and location could put a constrain on the ratio of formation channels, and, hence, the stability of mass transfer.

%%%%%%%%%%%%%%%%%%%%%%%%%%%%%%%%%%%%%%%%%%%%%%%%%%%%

\section{Metallicity Specific Star Formation Rate}
\label{MSSFR}

We divide the calculation of the \ac{MSSFR} into two independent factors, the \ac{SFR} and the metallicity distribution:
\begin{equation}
\label{eq:MSSFR}
 \dfrac{d^3 M_\mathrm{SFR}}{d t_s d V_c d Z}(z) = \dfrac{d^2 M_\mathrm{SFR}}{d t_s dV_c}(z) \times \frac{d P}{d Z} (z).
\end{equation}
In practice, the \ac{SFR} and the metallicity distribution may be correlated (see for example \citealt{furlong2015evolution});  however, decoupling the \ac{SFR} and the metallicity distribution is a convenient simplifying assumption that yields sufficient degrees of freedom given current observational constraints.

We discuss detailed models of the \ac{SFR} and metallicity distribution in Appendix~\ref{AppMSSFR}.  Here, we summarise the key approach to justify the shape of a phenomenological model that can be used for future inference.  We highlight a particular choice of the model parameters that, coupled with our default binary evolution model, produce a good match to data from the first two observing runs of the advanced detector network (see section \ref{GWdetections}).

\begin{figure}
 \includegraphics[width=\columnwidth]{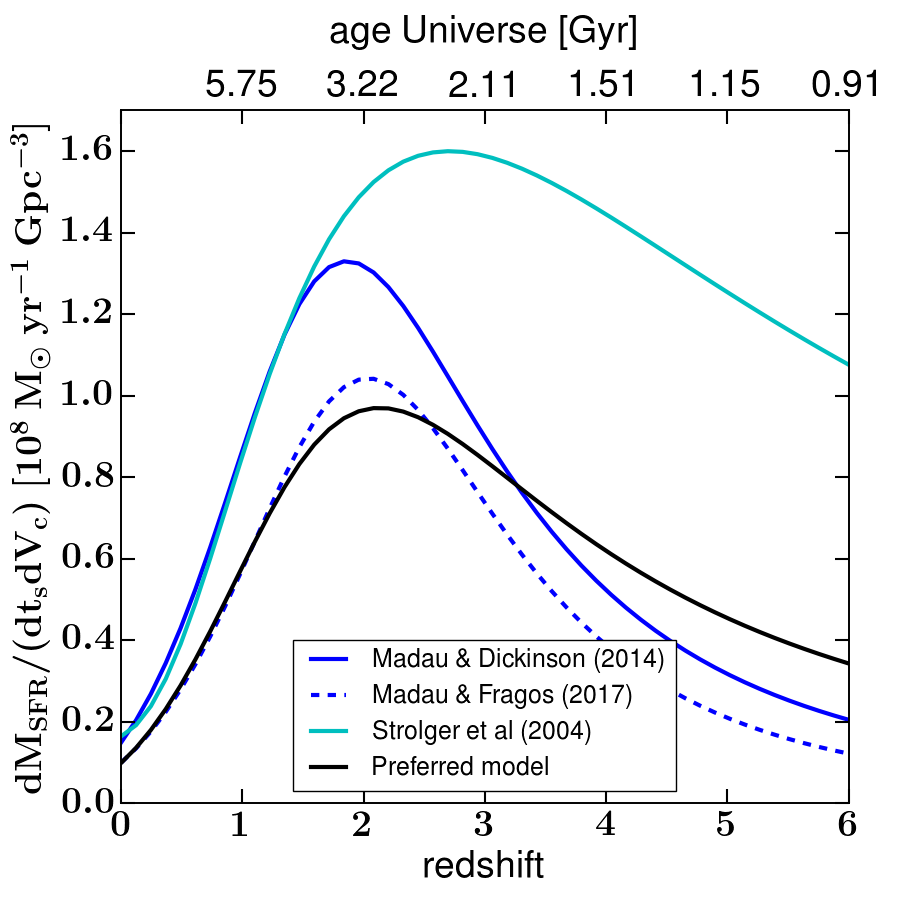}
    \caption{The star formation rate: mass per unit comoving volume per unit time as a function of redshift.
The distributions shapes are similar up to a redshift of 1.5. The star formation rate estimate of 
\citet{madau2014cosmic} peaks slightly earlier, at a redshift of two, and then decreases
steeply. \citet{strolger2004hubble} used an additional extinction correction and recover
a higher star formation rate at higher redshifts. \citet{2017ApJ...840...39M} assume a slightly different IMF resulting in a lower normalisation. We also include our preferred phenomenological model for the star formation rate history.}
    \label{fig:SFRplot}
\end{figure}

Figure~\ref{fig:SFRplot} illustrates the \ac{SFR} models we use.  All models agree well at low redshift, $z \lesssim 2$, other than differences in calibration due to the assumed initial mass function \citep{2017ApJ...840...39M}.  At higher redshift, \cite{strolger2004hubble} assume greater extinction and find a higher rate of star formation than \citet{madau2014cosmic}.  We follow the functional form of \citet{madau2014cosmic} in our phenomenological model:
\begin{equation}
  \frac{d^2 M_\mathrm{SFR}}{dt_sdV_c} = a \frac{(1+z)^{b}}{1+[(1+z)/c]^{d}}  \ \ \rm M_{\odot}\ year^{-1}\ Mpc^{-3}.
\end{equation}
The entire parameter space would be 4-dimensional, but we find that all of these \ac{SFR} prescriptions can be reasonably reproduced by setting $b=2.77$, c=$2.9$ and letting $a,d$ vary in the intervals [0.01--0.015] and [3.6--5.6], respectively.   In section \ref{GWdetections}, we show that $a=0.01$ and $d=4.7$ yield a good match to gravitational-wave observations when coupled with the metallicity distribution model discussed below and our default binary evolution model.

\begin{figure}
\includegraphics[width=\columnwidth]{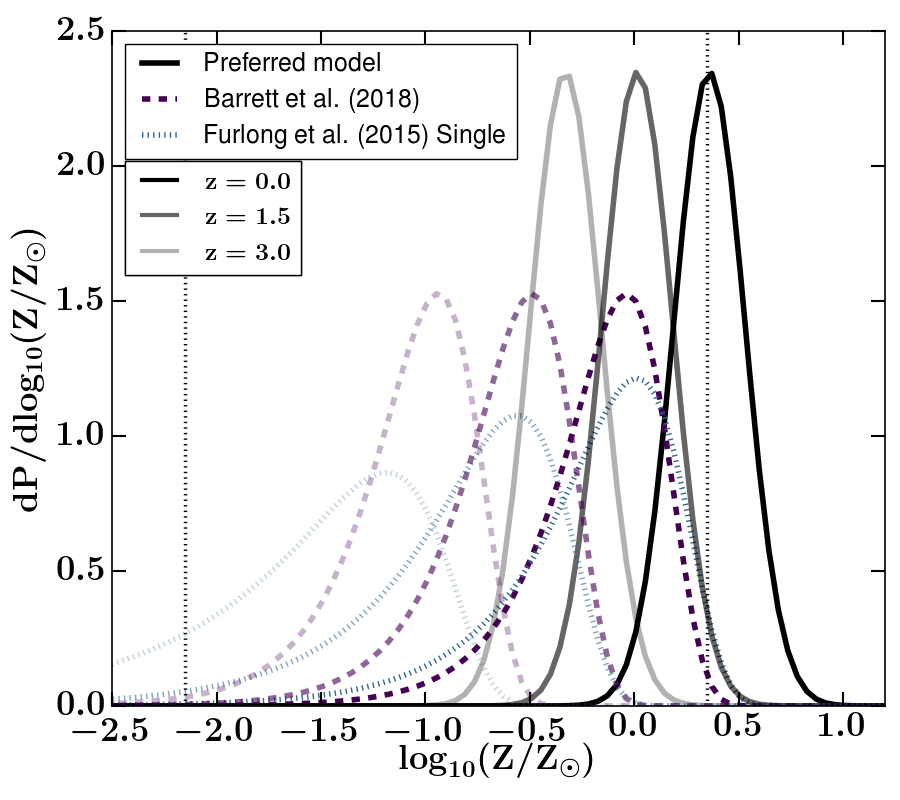}
\caption{The star formation metallicity distribution. The shades (dark to light) denote the redshifts 0, 1.5, and 3. Our previous model of \citet{2018JimInference} convolves the \ac{MZ}-relation of \citet{langer2006collapsar} with a redshift-independent \ac{GSMF} of \citet{panter2004mass} (purple dashed line). The blue dotted line instead uses the \ac{GSMF} by \citet{furlong2015evolution}. We also include our preferred model for the metallicity distribution of star formation (black solid).
The vertical dotted lines denote the limits of our metallicity grid; portions of the distribution extending beyond these limits are included in the edge bins when integrating over metallicity.}
\label{fig:ZvsPDF}
\end{figure}

The metallicity density function at each redshift is typically obtained by convolving a \ac{GSMF} with a \ac{MZ} relation.  Both of these are subject to significant uncertainties, and we describe several \ac{GSMF} fits \citep{panter2004mass, furlong2015evolution} and \ac{MZ} relations \citep{savaglio2005gemini, langer2006collapsar, ma2015origin} in 
Appendix~\ref{AppMSSFR}.  We show the metallicity distribution at several redshifts from a combination of some of these predictions in figure \ref{fig:ZvsPDF}.  This figure also shows our fiducial model -- a log-normal distribution in metallicity
\begin{equation}
\label{eq:lognormal}
\frac{dP}{dZ} (z)= \frac{1}{Z \sigma \sqrt{2\pi}} e^{-\frac{(\ln(Z)-\mu(z))^2 }{2\sigma^2}},
\end{equation}
with redshift-independent standard deviation $\sigma$ in $\ln(Z)$ space around a redshift-dependent mean $\mu$ of $\ln(Z)$ given by
\begin{equation}
 \langle Z \rangle = e^{(\mu + \frac{\sigma^2}{2})}.
\end{equation}
We follow \citet{langer2006collapsar} in parametrising mean metallicity as 
\begin{equation}
 \langle Z (z) \rangle = Z_{0} 10^{\alpha  z},
\end{equation}
where $Z_{0}$ is the mean metallicity at $z=0$ and the parameter $\alpha$ has negative values, yielding lower mean metallicity at higher redshifts.  Therefore the free parameters for the metallicity distribution are $Z_{0}$, $\alpha$ and $\sigma$.  We show in section \ref{GWdetections} that $Z_0=0.035$, $\alpha=-0.23$ and $\sigma=0.39$ yield a good match to gravitational-wave observations when coupled with our other assumptions; this preferred model has a similar shape to the metallicity distribution inferred by \citet{Rafelski:2012} from measurements of damped Lyman $\alpha$ galaxies.

%%%%%%%%%%%%%%%%%%%%%%%%%%%%%%%%%%%%%%%%%

\section{\ac{DCO} Mergers}
\label{CosmicDCOrate}

In this section we focus on the rate and mass distribution of \ac{DCO} mergers as a function of redshift. We convolve the \ac{DCO} population formed at each redshift (section \ref{COMPAS})
with the \ac{MSSFR} (section \ref{MSSFR}), incorporating the delay time distribution according to equation \ref{eq:intrinsic}. We do not yet take into account any selection effects.
We find that the choice of \ac{MSSFR} affects the total merger rate as a function of redshift, the relative rate between different types of \ac{DCO}, and the mass distribution. Additionally we show that our predicted distributions do not match the priors used by \citet{abbott2016binary,2018arXiv181112940T,LVC2018O2} for inference on gravitational-wave signals.

\subsection{Rate and Redshift of Cosmic \acp{DCO} Mergers}
\label{mergerRate}

\begin{figure}
\includegraphics[width=\columnwidth]{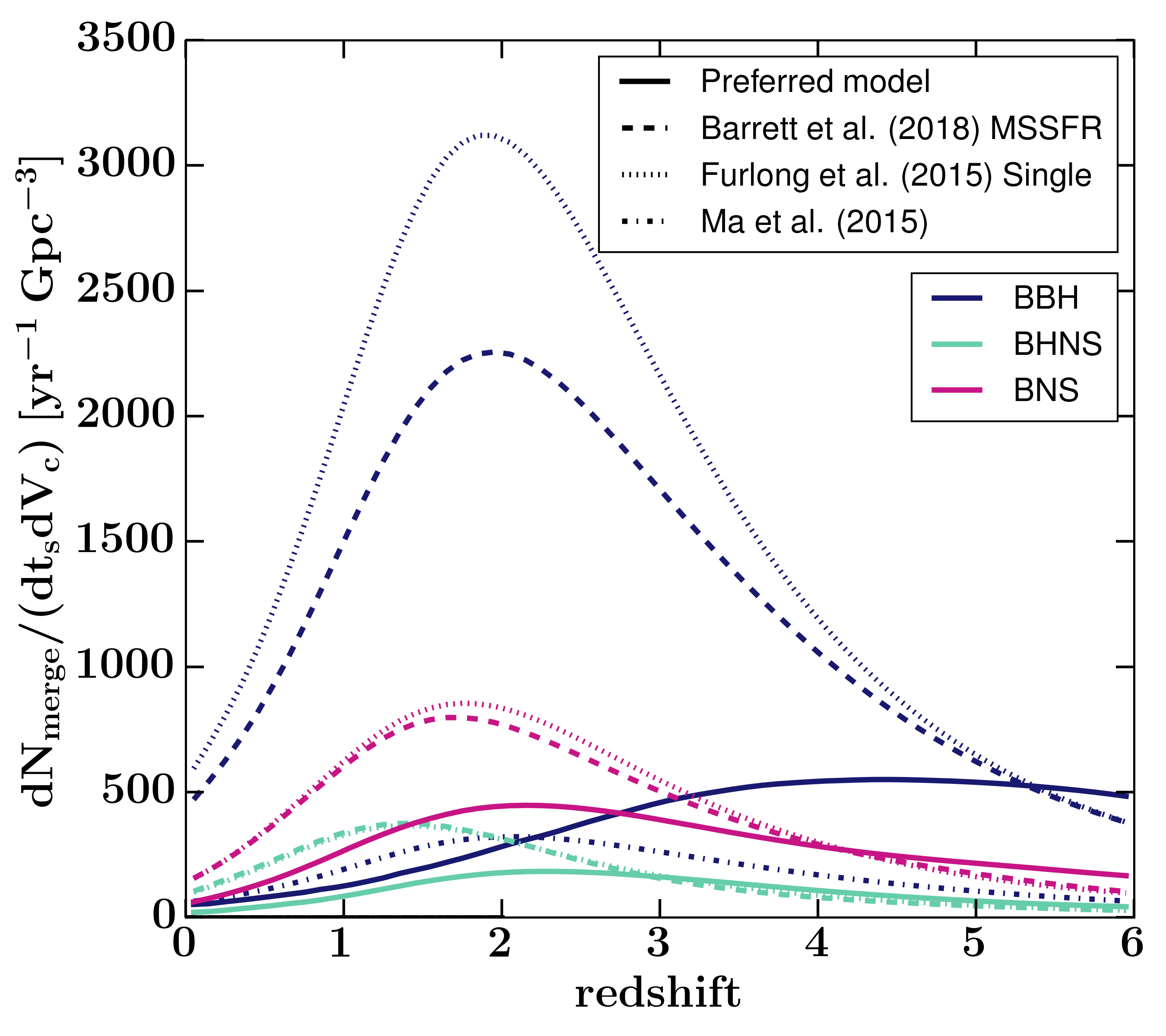}
      \caption{The intrinsic rate of \ac{DCO} mergers per cubic Gpc per year. The colours denote different \ac{DCO} types: \acp{BBH} in dark blue, \acp{BHNS} in mint, and \acp{BNS} in pink. The solid line is our preferred phenomenological model. The dashed line is the default model of \citet{2018JimInference}, which combines the \ac{SFR} of \citet{madau2014cosmic}, the \ac{MZ}-relation of \citet{langer2006collapsar}, and the redshift-independent \ac{GSMF} of \citet{panter2004mass}. The dotted line replaces the latter with the redshift-dependent single Schechter \ac{GSMF} of \citet{furlong2015evolution}. For clarity we only show the \ac{BBH} distribution for the \ac{MZ}-relation by \citet{ma2015origin} (dot-dashed).}
    \label{fig:TotalMergerRateRedshift}
\end{figure}

Figure \ref{fig:TotalMergerRateRedshift} shows the intrinsic rate of \ac{DCO} mergers as a function of redshift for a few \ac{MSSFR} combinations. In our preferred \ac{MSSFR} model, the merger rate at redshift $z=0$ is 49, 57, 20~$\rm Gpc^{-3}\ yr^{-1}$ for \acp{BBH}, \acp{BNS}, and \acp{BHNS}, respectively.  These rates are lower than the other models considered in section \ref{MSSFR}, including our previous \ac{MSSFR} model \citep{2018JimInference}.   The main reason is that our new preferred model favours higher metallicities at low redshifts (see figure \ref{fig:ZvsPDF}).  This suppresses the yield of \acp{BBH} and shifts their peak merger rate to higher redshifts which is in principle measurable with future gravitational-wave observations \citep{2018ApJ...863L..41F,2018arXiv180800901V}.  This is consistent with results from other \ac{MSSFR} models: the variation with the \citet{ma2015origin} \ac{MZ} relation, which has a higher average metallicity than the \citet{langer2006collapsar} \ac{MZ} relation used in \citet{2018JimInference}, and consequently yields lower \acp{DCO} merger rates; while the \citet{furlong2015evolution} redshift-dependent \ac{GSMF}, which allows for more low-mass galaxies than the \citet{panter2004mass} \ac{GSMF} assumed in \citet{2018JimInference}, with correspondingly lower metallicities, yields higher \ac{DCO} merger rates.

The effect of different metallicity distributions is smaller for the rates of \acp{BNS} and \acp{BHNS}, since their yield is less metallicity-dependent. We note that the change in the \ac{MSSFR} affects not only the overall \ac{DCO} merger rate, but also the ratio between different merger rates of different \ac{DCO} types.

\subsection{Mass Distribution and Redshift of Cosmic \ac{DCO} Mergers }\label{massDCO}

\begin{figure}
\includegraphics[width=\columnwidth]{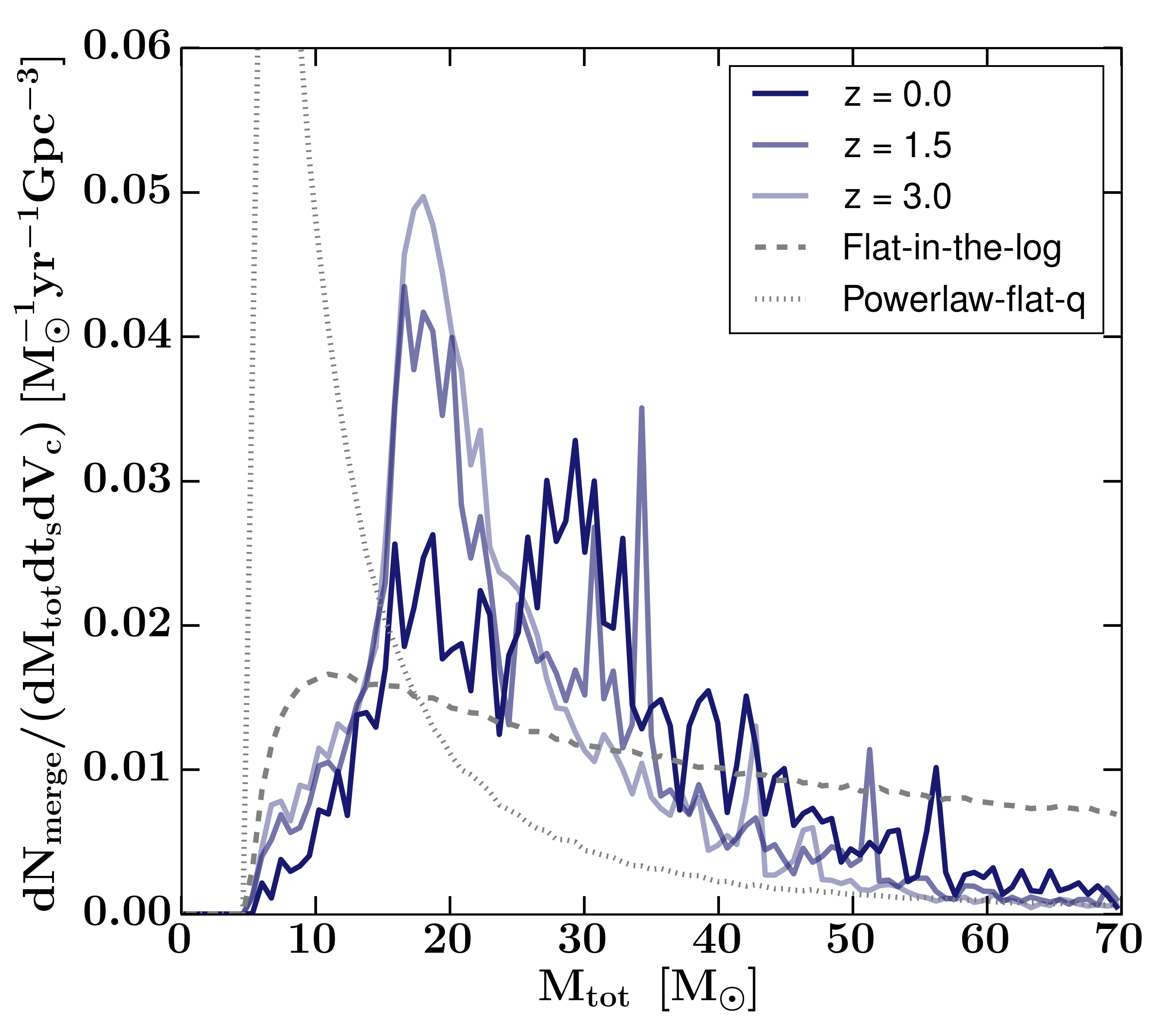}
      \caption{The normalised total mass distribution of \ac{BBH} mergers at redshifts 0, 1.5, and 3 (shaded dark to light) for our preferred log model. The narrow significant spikes above 30~$\rm M_{\odot}$ relate to the \ac{LBV} systems (for more details see appendix~\ref{AppRemnant}).
For comparison, the dotted curve indicates the total mass distribution assuming that the more massive \ac{BH} is sampled from a power law with index of -2.3 paired with a companion drawn from a flat mass ratio distribution. The dashed curve is a total mass distribution where both \ac{BH} masses are sampled from a flat-in-the-log distribution. For the minimum mass we took 2.5~$\rm M_{\odot}$ given that we have not introduced a mass gap. For the maximum mass we took 100~$\rm M_{\odot}$. In our simulations such high \ac{BH} masses are rare, but possible. Similar priors have been used in rate estimate studies such as \citet{abbott2016binary} and \citet{2018arXiv181112940T}.}
    \label{fig:MtotRedshift}
\end{figure}

Figure \ref{fig:MtotRedshift} shows the normalised total mass distribution of \ac{BBH} mergers at several redshifts for our preferred model \ac{MSSFR} model.  This is due to the convolution of the redshift dependence of the \ac{MSSFR} with the delay time distribution.  There is a significant contribution to low-redshift mergers from \acp{DCO} that formed at low metallicity and high redshift, with long delay times (see Fig.\ref{fig:tDelay}).   These low-metallicity systems give rise to high-mass \ac{BBH} mergers (see Fig.~\ref{fig:Mtot}). In fact, there is a greater tail of high-mass \acp{DCO} merging at redshift $z=0$ than at higher redshifts in figure \ref{fig:MtotRedshift} (see also \citet{2015DominikIIIGWrates, belczynski2016first}).

\begin{figure}
\includegraphics[width=\columnwidth]{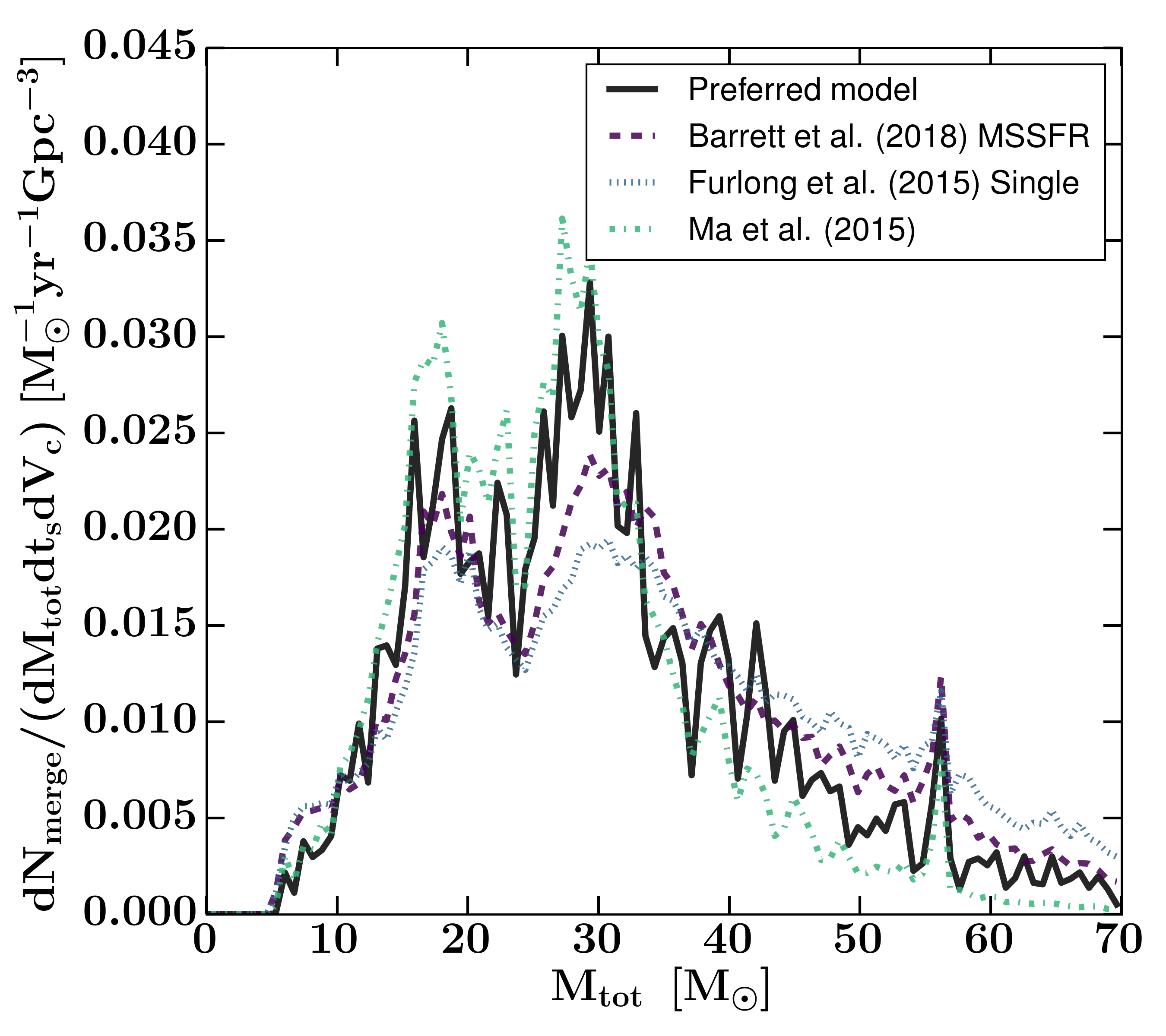}
      \caption{The normalised total mass distribution of \ac{BBH} mergers at redshift $z=0$. The \ac{MSSFR} models are the same as in figure \ref{fig:TotalMergerRateRedshift}.}
    \label{fig:MtotRedshift0}
\end{figure}

The mass distribution is sensitive to the metallicity of formation, and therefore depends on the assumed \ac{MSSFR} prescription.  We show the impact of the \ac{MSSFR} on the mass distribution of \acp{DCO} merging at redshift $z=0$ in figure \ref{fig:MtotRedshift0}.  As with the \ac{BBH} merger rate discussed in section \ref{mergerRate}, \ac{MSSFR} models with lower metallicity (our previous  model in \citet{2018JimInference}, especially with the \citet{furlong2015evolution} \ac{GSMF} variation) show enhanced high-mass tails relative to \ac{MSSFR} models with higher metallicity  (our preferred model )or reduced high-redshift, low-metallicity \ac{SFR} (the \citet{ma2015origin} \ac{MZ} relation combined with the \ac{SFR} of \citet{madau2014cosmic}, the redshift-independent \ac{GSMF} of \citet{panter2004mass}).   The peaks at high masses in figures~\ref{fig:MtotRedshift} and \ref{fig:MtotRedshift0} are due to mass loss prescriptions, particularly \ac{LBV} winds (for more details see Appendix~\ref{AppRemnant}). These depend on metallicity, hence the prominence of the peaks varies depending on the \ac{MSSFR} prescription.

%%%%%%%%%%%%%%%%%%%%%%%%%%%%%%%%%%%%%%%%%

\subsection{Priors and Rate Estimates}
\label{subsec:priors_rates}

The \ac{DCO} merger rate inferred from gravitational-wave observations is sensitive to the assumed mass distribution  \citep{2018arXiv181112940T, LVC2018O2}.  We show the mass priors assumed by \citet{abbott2016binary} and \citet{2018arXiv181112940T} in figure \ref{fig:MtotRedshift}; it is clear that these are inconsistent with our predicted mass distribution.   \citet{2018arXiv181112940T} account for uncertainties in the shape of the \ac{BBH} mass distribution by varying the slope of a power law distribution.  However, as we show in Fig.~\ref{fig:MtotRedshift}, the mass distribution of \acp{BBH} might be more complex than a simple power law, and is furthermore a function of redshift, along with the merger rate itself.  Therefore, \ac{DCO} merger rates and mass distributions inferred from simple priors or phenomenological models should be treated with caution.

The complex dependence of the mass distribution of merging \acp{DCO} on both the binary evolution model \citep[e.g.][]{2013DominikIIRates, 2017MapelliCosmicBH, stevenson2017formation} and the \ac{MSSFR} (this study), and the variation in the mass distribution and merger rate with redshift, makes it challenging to propose alternative priors.  Therefore, it is preferable to apply selection effects to the model population in order to compare model predictions against observations.  This is the approach we take in the next section.

%%%%%%%%%%%%%%%%%%%%%%%%%%%%%%%%%%%%%%%%%

\section{Gravitational-Wave Detections}
\label{GWdetections}

This section focuses on the effect of the \ac{MSSFR} on the predicted rates and mass distributions of detectable \ac{DCO} mergers. We evaluate these using Eq.~\ref{eq:detections}.  We predict the total rate of detectable \ac{DCO} mergers as a function of redshift and describe the mass distribution of \ac{BBH} mergers. We carry out a Bayesian model comparison of different \ac{MSSFR} prescriptions, taking into account both the number and the mass estimates of the 10 \ac{BBH} mergers detected during the first and second observing runs of \ac{aLIGO} \citep{abbott2016binary, LVC2018O2}.  We do not include in our analysis the 6 additional \ac{BBH} candidates found in the same data set by \citet{2019arXiv190407214V} with an independent search pipeline and somewhat different data quality choices.

\subsection{Selection Effects}
\label{selection}

For the selection effects we use the same method as described in \citet{2018JimInference}. We use a single detector \ac{SNR} threshold of 8 \citep{aasi2016prospects}, above which we assume that gravitational waves from the merger are detectable.  To evaluate the \ac{SNR} for a given \ac{DCO} system, we compute the waveforms for the appropriate masses using a combination of IMRPhenomPv2 \citep{Hannam:2013oca, Husa:2015iqa, Khan:2015jqa} and SEOBNRv3 \citep{Pan:2013rra, Babak:2016tgq}.  We approximate the sensitivity of the second observing run \citep{LVC2018O2} to be similar to the first observing run \citep{abbott2016binary}.  The fraction of systems with \ac{SNR} above the threshold of 8 at a given distance (redshift), after sampling over the sky location and orientation of the binary \citep{FinnSkyAverage}, yields the detection probability $P_\textrm{det} (M_\textrm{chirp}, D_L)$.

\subsection{Rate and Redshift of Gravitational-Wave Detections}
\label{subsec:rate_redshift_detections}

\begin{figure}
\includegraphics[width=\columnwidth]{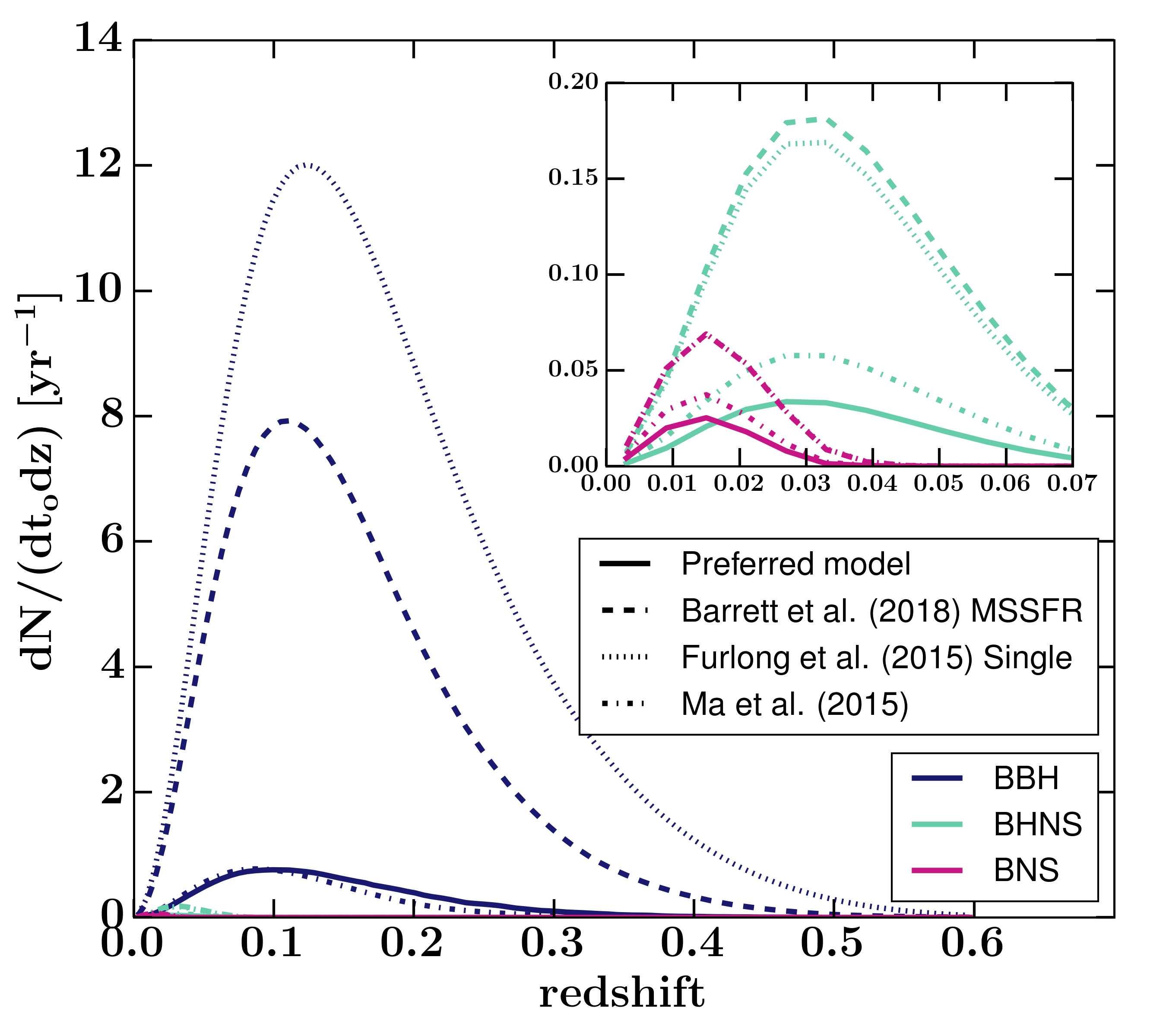}
      \caption{The number of \ac{DCO} mergers per year per unit redshift at the sensitivity of the first two advanced detector observing runs.   The top-right panel is an enlargement with the same axes to focus in on the merger rates for \acp{BHNS} and \acp{BNS}.  The \ac{MSSFR} models are the same as in Fig.~\ref{fig:TotalMergerRateRedshift}.}
    \label{fig:TotalMergerRateRedshift2}
\end{figure}

The rate of detectable \ac{DCO} mergers depends on the underlying merger rate, which increases up to redshift $z \sim 2$  (Fig.~\ref{fig:TotalMergerRateRedshift}).  However, the detection probability drops off at higher redshift.  These competing effects mean that the detection rate of \ac{BBH} mergers this results in a peak rate at a redshift between 0.1--0.15 depending on the \ac{MSSFR} model at the sensitivity of the first two observing runs.  This is shown in Fig.~\ref{fig:TotalMergerRateRedshift2}.  Note that this figure displays the number of detections per unit redshift per unit observer time, rather than per unit volume per unit source time as in Fig.~\ref{fig:TotalMergerRateRedshift} (see Eq.~\ref{eq:detections} for the additional factors of $dV_c/dz / (1+z)$).  Because mergers involving less massive \acp{NS} cannot be observed as far as \ac{BBH} mergers, detection rates of \ac{BHNS} and \ac{BNS} mergers per unit redshift peak at $z \approx 0.03$ and $z \approx 0.015$, respectively. As discussed in Sec.~{\ref{mergerRate}, sensitivity of the detection rate to \ac{MSSFR} variations tracks the sensitivity of the \ac{DCO} formation rate to metallicity (see Fig.~\ref{fig:yield}).

\begin{table*}
\begin{tabular}{|l |l | l| llllll}
\multicolumn{3}{|c|}{Variation MSSFR} & \multicolumn{3}{|c|}{ Detection rate [yr$^{-1}$]} & \multicolumn{3}{|c|}{Likelihoods ($\log_{10}$)} \\
SFR & MZ & GSMF & $\rm BBH $ &   $\rm BHNS $  &$\rm BNS$ &   $\rm \mathcal{L}_{M_{chirp}}$ & $\rm \mathcal{L}_{R}$ & $\rm \mathcal{L}_{tot}$\\ \hline \hline
\multicolumn{3}{|c|}{Preferred model} & 22.15  & 0.23  & 0.08 & -32.32  & -0.90 & -33.22 \\
\hline
Madau et al. & Ma et al. (2004) & 1 &18.43 & 0.4 & 0.11 & -33.9 & -0.97 & -34.87 \\
 & & 2 &94.35 & 0.51 & 0.12 & -32.42 & -8.86 & -41.28 \\
 & & 3 &113.92 & 0.52 & 0.13 & -32.48 & -11.9 & -44.38 \\\cline{2-9}
 & Langer et al.  &1 &247.22 & 1.28 & 0.22 & -32.24 & -34.85 & -67.09 \\
 &  &2 &441.08 & 1.19 & 0.22 & -32.61 & -70.6 & -103.21 \\
 &  &3 &492.27 & 1.25 & 0.23 & -32.77 & -80.23 & -113.0 \\\cline{2-9}
 & Langer et al., offset  &1 &28.72 & 0.23 & 0.09 & -32.3 & -1.07 & -33.38 \\
 &  &2 &120.3 & 0.35 & 0.11 & -32.68 & -12.93 & -45.61 \\
 &  &3 &148.74 & 0.35 & 0.11 & -32.87 & -17.62 & -50.49 \\\hline
Strolger et al. & Ma et al. (2004) & 1 &32.93 & 0.52 & 0.12 & -33.82 & -1.31 & -35.13 \\
 & &2 &203.93 & 0.6 & 0.14 & -32.81 & -27.14 & -59.95 \\
 & &3 &208.21 & 0.61 & 0.14 & -32.65 & -27.9 & -60.54 \\\cline{2-9}
 & Langer et al. &1 &406.39 & 1.28 & 0.23 & -32.44 & -64.11 & -96.55 \\
 &  &2 &659.25 & 1.19 & 0.24 & -32.98 & -111.92 & -144.9 \\
 &  &3 &710.91 & 1.25 & 0.24 & -33.09 & -121.79 & -154.87 \\\cline{2-9}
 & Langer et al., offset &1 &89.79 & 0.33 & 0.11 & -32.46 & -8.18 & -40.63 \\
 &  &2 &267.34 & 0.43 & 0.12 & -33.2 & -38.48 & -71.68 \\
 &  &3 &292.76 & 0.43 & 0.12 & -33.22 & -43.1 & -76.33 \\ \hline
\end{tabular}
\caption{Rate estimates and likelihoods per \ac{MSSFR} variation. The numbers in the column \ac{GSMF} refer to 1=\citet{panter2004mass}, 2=\citet{furlong2015evolution} (single Schechter function), 3=\citet{furlong2015evolution} (double Schechter function). The detection rates are estimated for a year of coincident observing with the sensitivity of the first observing run of \ac{aLIGO}. The likelihoods account for \ac{BBH} detections during the first and second observing runs, assuming the same sensitivity \citep{abbott2016binary, LVC2018O2}. The total log likelihood $\rm \mathcal{L}_{tot}$ is the sum of the log likelihoods of the chirp-mass distribution $\rm \mathcal{L}_{M_{chirp}}$ and the rate $\rm \mathcal{L}_{R}$.}
\label{tab:mainResult}
\end{table*}

Table~\ref{tab:mainResult} shows the observed rate per \ac{DCO} type per year. The combined observing time of the first two observing runs is about 166 days: 48 days of coincident data for the first and 118 days for the second observing run \citep{abbott2016binary, LVC2018O2}.
Thus, 10 detections translate to an observed detection rate of 22 \ac{BBH} mergers per year. Most of our variations significantly overestimate the observed rate.  As previously mentioned, variations that favour a higher \ac{SFR} \citep{strolger2004hubble}, lower metallicities \citep{langer2006collapsar} or lower galaxy stellar masses \citep{furlong2015evolution} predict a higher detection rate. We find that by changing the \ac{MSSFR} alone we can vary the predicted rate of detectable \ac{BBH} mergers by more than an order of magnitude. 

All of the predictions for detectable \ac{BNS} mergers are lower than of one in four years of observing time, suggesting that GW170817 was a fortuitous event.  \ac{MSSFR} models with the highest rates predict more than one detectable \ac{BHNS} merger in one year observing time, however, these are generally inconsistent with observations in their \ac{BBH} merger rate predictions.

\subsection{Mass Distribution of Detectable \ac{BBH} Mergers}
\label{subsec:mass_dist_detectable_BBH}

\begin{figure}
\includegraphics[width=\columnwidth]{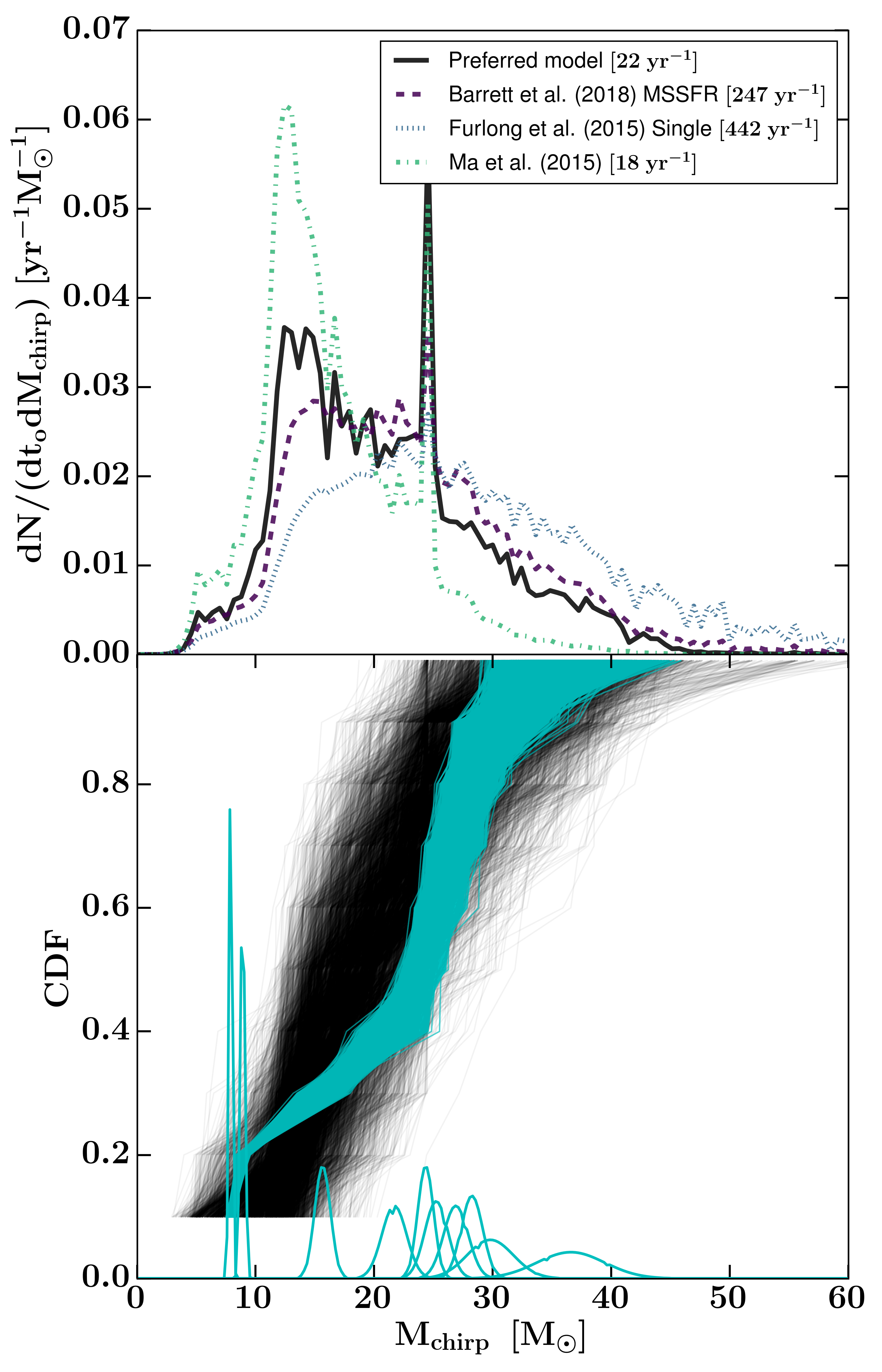}
      \caption{Top panel: Predicted chirp-mass distribution of \ac{BBH} mergers detectable at the sensitivity of the first two observing runs. The masses are in the source reference frame. The \ac{MSSFR} models are the same as in Fig.~\ref{fig:TotalMergerRateRedshift}.  The numbers in the label are the predicted annual detection rate. Bottom panel: Approximate Gaussian posteriors (see appendix \ref{Bayes}) for \ac{BBH} mergers detected during the first and second observing runs \citep{LVC2018O2}, from left to right: GW170608, GW151226, GW151012, GW170104, GW170814, GW170809, GW170818, GW150914, GW170823, GW170729. The cyan area shows randomly sampled cumulative density functions from the posteriors, indicating the spread due to the measurement uncertainty.  The black lines are cumulative density functions when 10 events are randomly drawn from the preferred model.}
    \label{fig:MtotRedshift01}
\end{figure}

The top panel of Fig.~\ref{fig:MtotRedshift01} shows the predicted chirp-mass distributions of detectable \ac{BBH} mergers for several \ac{MSSFR} variations.  We use the chirp masses of the \acp{BBH} mergers here since these are typically better observationally constrained than the total masses.

Mergers of more massive \acp{DCO} emit louder gravitational-wave signal that can be detected to greater distances.  Therefore, the mass distribution of detectable \acp{BBH} is shifted to higher masses relative to the intrinsic mass distribution of Fig.~\ref{fig:MtotRedshift0}.  The impact of \ac{MSSFR} variations on the shape of the distribution follows the discussion in Sec.~\ref{massDCO}. However, the selection effects emphasise the peak due to the \ac{LBV} winds at chirp masses around $\rm 25\ M_{\odot}$ (see also appendix~\ref{AppRemnant}).  Although the 'delayed' remnant mass model of \citet{2012FryerRemnants} used in our simulations does not enforce a mass gap between \ac{NS} and \ac{BH} masses, we find that low-mass merging \acp{BBH} are very rare, especially after selection effects are applied.   We do not expect significant numbers of detections in the mass gap for any of the \ac{MSSFR} variations.

The bottom panel of Fig.~\ref{fig:MtotRedshift01} shows that our preferred \ac{MSSFR} model predicts a chirp mass distribution of detectable \ac{BBH} mergers that is consistent with the detection from the first two advanced detector observing runs.  Approximate Gaussian posteriors (see appendix~\ref{Bayes}) for the ten detections are shown in cyan at the bottom of the plot.  We construct observed CDFs by taking a random sample from each of these ten posteriors.  The set of cyan curves  indicates the range of observed CDFs consistent with measurement uncertainty.  Meanwhile, each black curve represents a CDF constructed by sampling from the predicted distribution of detectable \ac{BBH} events under the preferred \ac{MSSFR} model.  The visual consistency between the black and cyan regions indicates a successful graphical predictive check of the model.

\subsection{Bayesian Comparison of MSSFR Models}
\label{subsec:bayesian}

We showed that the choice of the \ac{MSSFR} affects both the detectable rates and mass distributions of \ac{DCO} mergers. Here we quantitatively compare these models against observations during the first and second observing run \citep{LVC2018O2}. We consider the total rate of events and the relatively well-measured chirp masses.  We do not consider other properties such as relatively poorly measured mass ratios or source redshifts given the narrow range of redshifts reached to date.  \citet{simone} compare a possible model for \ac{BBH} spins evolving through channel I (see section~\ref{subsec:channels}) using the COMPAS data presented here against observations.  In this analysis (as in \citet{2018JimInference}), the total log likelihood $\mathcal{L}_\mathrm{tot}$ is the sum of the rate log likelihood $\mathcal{L}_\mathrm{R}$ and the likelihood of the normalised chirp-mass distribution $\mathcal{L}_\mathrm{M_{chirp}}$:
\begin{equation}
\mathcal{L}_\mathrm{tot} = \mathcal{L}_\mathrm{M_{chirp}}  + \mathcal{L}_\mathrm{R}.
\end{equation}
The rate likelihood assumes a Poisson distribution where the \ac{MSSFR} model gives the expected number of detections over the duration of the first two observing runs. The chirp-mass likelihood is the product over the ten events of the probabilities of making individual detections given the predicted chrip-mass distribution (see appendix~\ref{Bayes}).   A difference of 1 in log likelihoods, corresponding to a factor of 10 in the likelihoods, implies that the higher-likelihood model is preferred over the lower-likelihood model by a factor of 10 (i.e., has an odds ratio of $10:1$, assuming both models are equally probable a priori).    Table~\ref{tab:mainResult} shows the total likelihoods for the pessimistic common-envelope assumptions. A longer list of variations, including the optimistic common-envelope assumption, can be found in tables~\ref{allLikelihoods}, \ref{allRates}.

The rate likelihoods differ significantly given our range in rate estimates. Many of the \ac{MSSFR} models greatly overestimate the rates and are strongly disfavoured under the assumed model of  binary evolution.
Meanwhile, despite the visual difference in the shape of the chirp-mass distribution (see Fig.~\ref{fig:MtotRedshift01}), the difference in the chirp-mass likelihoods is small. More detections will make it possible to jointly explore \ac{MSSFR} and evolutionary models using the observed chirp-mass distributions \citep{2018JimInference}.
Given our binary evolution model, higher star-formation metallicities
at low redshifts are preferred to match the observed \ac{BBH} rate and chirp-mass distribution. 

In section \ref{MSSFR} we introduced a 5-parameter phenomenological model of the \ac{MSSFR}.  With suitable parameter choices this generic model can match all of the detailed models considered here, while providing the convenience of a continuous, smooth parametrisation that is useful for inference.  We also introduced a particular choice of these 5 parameters -- our preferred model -- that yields a good match to both the number of \ac{BBH} mergers detected during the first two observing runs (10.06 predicted vs.~10 observed) and their chirp mass distribution (section \ref{subsec:mass_dist_detectable_BBH} and the bottom panel of Fig.~\ref{fig:MtotRedshift01}).  As table \ref{tab:mainResult} shows, this preferred model also yields the highest likelihood among all considered models. This preferred model favours a \ac{SFR} similar to \citet{2017ApJ...840...39M}, which includes the contribution from stars in binaries. However, we do favour a higher \ac{SFR} at high redshifts, where metallicity is lower, to enhance the fraction of massive \ac{BBH} merger events.  We caution, however, that the \ac{MSSFR} parameters in the preferred model are chosen ad hoc, with some `Fingerspitzengef\"{u}hl'.  Future analyses should jointly infer the parameters of the \ac{MSSFR} and parameters describing the binary evolution model, using gravitational waves and other observational constraints.

%%%%%%%%%%%%%%%%%%%%%%%%%%%%%%%%%%%%%%%%%%%%%%%%%%%%%%%%%

\section{Conclusion and Discussion}
\label{discussion}

We showed that assuming different \ac{MSSFR} within observational constraints can vary the rate of \ac{BBH} mergers by more than an order of magnitude within a fixed stellar and binary evolution model\footnote{There is a further uncertainty from the definition of solar metallicity, see appendix \ref{subsec:solar_values}.} and affect the ratio between \ac{BBH} and \ac{BNS} detection rates.
This is comparable to the impact of uncertainties on evolutionary physics such as wind mass loss rates, conservativeness of mass transfer, the efficiency of common envelope evolution and \ac{BH} natal kicks \citep{2012DominikCEE,2018MNRAS.481.1908K,giacobbo2018progenitors}.

The sensitivity to \ac{MSSFR} is predominantly driven by the impact of metallicity on the yield of \acp{BBH} per unit star forming mass.  This is consistent with earlier findings \citep[e.g.,][]{2015DominikIIIGWrates, chruslinska2018influence}.  In particular, \citet{chruslinska2018influence} also find that a higher average metallicity is required in order to not over-predict the \ac{BBH} merger rate.

Here, we explored the impact of the \ac{MSSFR} while keeping the binary evolution model unchanged.  In practice, joint inference on stellar and binary physics and the \ac{MSSFR} is required to fully interpret observations \citep[e.g.,][]{chruslinska2018influence}. For example, (pulsational) \acp{PISN} (see for example \citet{2017WoosleyPPISN} and references therein) can prevent the formation of \acp{BH} with masses between around 50 and 130 M$_\odot$, i.e., with chirp masses between 45 and 115 M$_\odot$ for equal-mass binaries.  \citet{2018arXiv181112940T} find that existing gravitational-wave detections show evidence for a maximum black hole mass of around $\sim 45$\,M$_\odot$, consistent with population synthesis studies such as \citet{2016A&A...594A..97B, 2017MNRAS.470.4739S, 2019arXiv190402821S}. However in Fig.~\ref{fig:MtotRedshift01}, we show that it is possible to reproduce such a limit within the evolutionary model of this paper, which does not include pulsation \acp{PISN}, by choosing a suitable \ac{MSSFR} alone (\citet{madau2014cosmic, ma2015origin, langer2006collapsar}). A similar argument can be made for the presence or absence of a mass gap between \acp{NS} and \acp{BH}.  Beyond gravitational-wave observations, other observational constraints such as the epoch of reionisation \citep{2016MNRAS.456..485S}  and X-ray binaries \citep{2017ApJ...840...39M} can further help to lift the degeneracy between binary physics and the \ac{MSSFR}.

\begin{figure}
\includegraphics[width=\columnwidth]{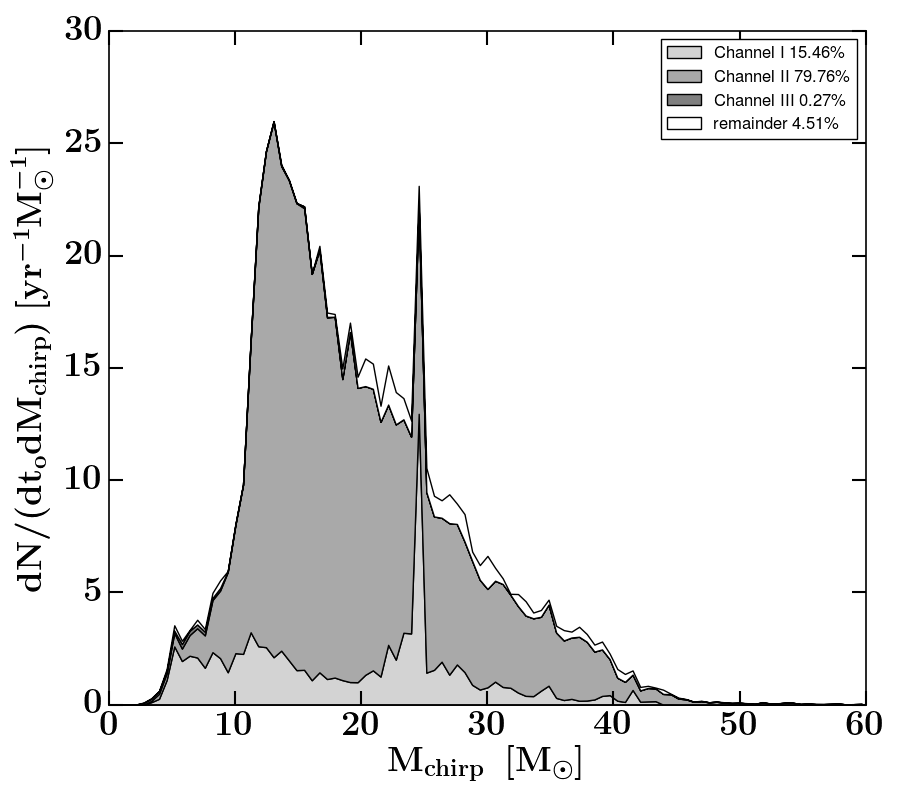}
      \caption{The predicted chirp mass distribution of detectable \ac{BBH} mergers at advanced detector design sensitivity within our preferred \ac{MSSFR} model, coloured by the \ac{BBH} formation channels.  Almost 80\% of the expected 380 detectable \acp{BBH} per year are formed through channel II (dynamically stable mass transfer) within our evolutionary model.}
    \label{fig:designChannels}
\end{figure}

We introduced a phenomenological description of the \ac{MSSFR} with 5 continuous parameters (section \ref{MSSFR}) to facilitate the joint exploration of the \ac{MSSFR} and parametrised evolutionary assumptions.  We also proposed a particular choice of the \ac{MSSFR} model parameters that represents a good match to the gravitational-wave detections made during the first two observing runs of advanced LIGO and Virgo.  Looking ahead, we can apply this preferred \ac{MSSFR} model to make predictions for the detection rate and chirp mass distribution at design sensitivity, shown in figure \ref{fig:designChannels}.  We predict 380 detections \ac{BBH} detections per year, or approximately one detection per day, within our default evolutionary model. 

Figure \ref{fig:designChannels} also highlights the importance of the dynamically stable mass transfer channel without a common envelope phase for the formation of detectable merging \acp{BBH}.  We find that channel II may be responsible for 80\% of all detected \ac{BBH} mergers.  This highlights the importance of mass transfer stability criteria, which merit further investigation.  Meanwhile, the narrow chirp mass spike at around 25 M$_\odot$ is due to the operation of LBV mass loss at a particular metallicity \citep[cf.][]{2015DominikIIIGWrates}.  While we expect that a finer metallicity grid or interpolation between metallicities would lead to a smoother chirp mass distribution, this again highlights the importance of highly uncertain LBV winds for these predictions \citep{Mennekens2014LBV}.  Finally, the sampling accuracy of predictions (e.g., the time delay distribution for \ac{BHNS} in Figure~\ref{fig:tDelay}) could be improved with more efficient importance sampling techniques \citep{2019arXiv190500910B}.

Our predictions suggest that approximately one thousand detections could be reached within a couple of years of operation of advanced detectors operating at design sensitivity.  \citet{2018JimInference} showed that this will be sufficient to constrain the binary evolutionary parameters to a fractional accuracy of a few percent.  Our phenomenological \ac{MSSFR} model can be incorporated into this hierarchical modelling framework to be enable joint inference on binary evolution and the cosmic history of star formation.

\section*{Acknowledgements}

C.~J.~N.~thanks the University of Birmingham for financial support,
C.~P.~L.~Berry, S.~Bavera and A.~Vecchio for helpful discussions, D.~J.~Stops for technical support, and of course above all T.~F.~Pauw for everything. A.~V.~G.~acknowledges support from Consejo Nacional de Ciencia y Tecnologia (CONACYT). S.~S.~is supported by the Australian Research Council Centre of Excellence for Gravitational Wave Discovery (OzGrav), through project number CE170100004.
S.~M.~is supported by STFC grant ST/M004090/1.
D.~Sz.~accepts funding from the Alexander von Humboldt Foundation.
This paper used the \texttt{Astropy} library \citep{Astropy1, Astropy2}, matplotlib \citep{Hunter:2007}, and numpy \citep{numpy}. 

%%%%%%%%%%%%%%%%%%%%%%%%%%%%%%%%%%%%%%%%%%%%%%%%%%



\bibliographystyle{mnras}
\bibliography{ms} % if your bibtex file is called example.bib

%\newpage 
%%%%%%%%%%%%%%%%%%%%%%%%%%%%%%%%%%%%%%%%%%%%%%%%%%

%%%%%%%%%%%%%%%%% APPENDICES %%%%%%%%%%%%%%%%%%%%%
\appendix

\section{Metallicity Specific Star Formation Rate}
\label{AppMSSFR}

\subsection{Cosmological Star Formation Rate - SFR}
\label{SFR}

We consider several prescriptions for the cosmological \ac{SFR} as a function of redshift $z$. The first is from \cite{madau2014cosmic}:
\begin{equation}
\label{eq:MadauDickinson} 
\dfrac{d^2 M_\mathrm{SFR}}{d t_s d V_c}(z) = 0.015 \frac{(1+z)^{2.7}}{1+[(1+z)/2.9]^{5.6}}\ \ \rm M_{\odot}\ yr^{-1}\ Mpc^{-3} .
\end{equation}
At higher redshifts the observations become more sensitive to extinction which is not exactly known. \cite{strolger2004hubble} construct a fit for the SFR using a different extinction correction, as
\begin{eqnarray}
\label{eq:sfr_strolger}
\dfrac{d^2 M_\mathrm{SFR}}{dt_s dV_c}(t) = 
0.182 \times \\
\left(t^{1.26}e^{-t/1.865} + 0.071\ e^{0.071(t - t_0)/1.865)}\right) \rm M_{\odot}\ yr^{-1}\ Mpc^{-3},
\nonumber
\end{eqnarray}
where $t(z)$ is the age of the Universe at redshift $z$ in Gyrs, and $t_0$ is the current age of the Universe, which they set to 13.47 Gyrs. These two \ac{SFR} models agree at low redshifts, $z \lesssim 2$, where both models peak; however, the model of \citet{strolger2004hubble} has a shallower drop off  at higher redshifts (see Fig.~\ref{fig:SFRplot}). Simulations so far have not independently constrained the \ac{SFR} at high redshifts. There are, for example, additional uncertainties such as the role of active galactic nuclei and feedback on the interstellar medium \citep{Taylor2015AGN}. 

\citet{2017ApJ...840...39M} use an updated \ac{SFR} compared to \citet{madau2014cosmic}. A key difference is assuming a broken power-law \ac{IMF} by \citet{2001MNRAS.322..231K} instead of the classic power-law by \citet{1955Salpeter}. This increases the relative number of massive stars and therefore lowers the overall \ac{SFR} normalisation by a factor of 0.66.  The shape of the \citet{2017ApJ...840...39M} and \citet{madau2014cosmic} \ac{SFR} models is similar, and we generally use the \citet{madau2014cosmic} prescription in our analysis.   However, we adjust the low-redshift normalisation of our preferred model to approximately match the more recent estimate of \citet{2017ApJ...840...39M}.

\subsection{Galaxy Stellar Mass to Metallicity - \ \ \ MZ-Relation}

As described in section \ref{MSSFR}, we can construct star-forming metallicity density functions by convolving the galaxy stellar mass distribution with the \ac{MZ} relation, which connects the galaxy stellar mass ($\rm M_{*}$) and metallicity.  We describe the \ac{MZ} relations considered in this work in this subsection, and the \acp{GSMF} in the next one.

\begin{figure}
  \includegraphics[width=\columnwidth]{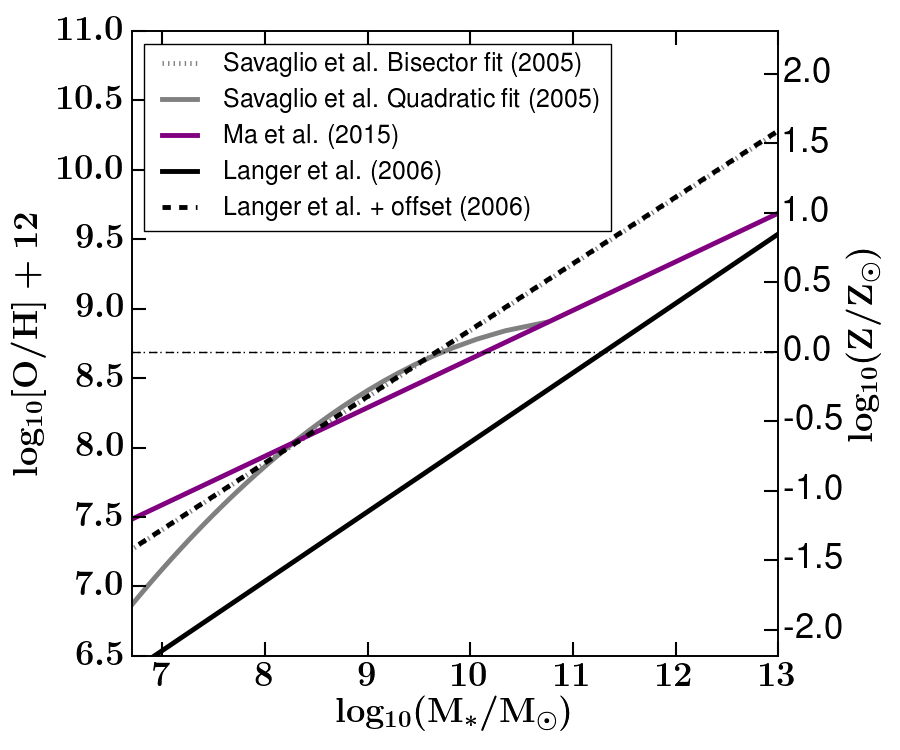}
   \caption{Comparison between different \ac{MZ} relations at a
   redshift of 0.7, at which the relation of \citet{savaglio2005gemini} was
   determined. We also show our introduced offset to \citet{langer2006collapsar} which
   overlaps with the bisector fit of \citet{savaglio2005gemini}. Note that the
   extrapolation of the quadratic fir of \citet{savaglio2005gemini} beyond their upper limit of $\rm
   \log_{10}(M_{*})=11$ results in a turnover. The dot-dashed horizontal line is our definition
   of the relation between solar metallicity and solar oxygen number density
   \citep{2009AsplundMetallicity}.}
    \label{fig:MZrelations}
\end{figure}

Stellar metallicities are assumed to match the metallicity of the interstellar gas of their surroundings at their birth.  Observations are typically given in terms of the ratio of the number density of oxygen and hydrogen in the gas, generally written as $\rm \log_{10}[O/H]+12$.  Conversions to metallicity depend on the assumed solar abundances.  In this study we define the solar metallicity mass fraction as $Z_{\odot}=0.0142$ and the solar oxygen abundance of $\log_{10} [\mathrm{O/H}]_{\odot}+12=8.69$ based on \citet{2009AsplundMetallicity}, but see appendix \ref{subsec:solar_values}. \citet{ma2015origin} discuss some of the uncertainties in the slopes and offsets in the \ac{MZ} relation, including the use of different observational samples or metallicity diagnostics, or the use of different simulation resolutions and feedback mechanisms in theoretical models \citep[e.g.,][]{Taylor2015AGN}.

In \cite{2018JimInference} we used the prescriptions of \citet{langer2006collapsar}, who in turn use a \ac{MZ} relation from \cite{savaglio2005gemini}. This \ac{MZ} relation is derived from a fit of 56 galaxies in the Gemini Deep Deep Survey with a mean redshift of around 0.7. \citet{savaglio2005gemini} provide a quadratic and linear bisector fit, the latter being
\begin{equation}
\label{eq:Savaglio}
\log_{10}[\mathrm{O/H}]+12 = 0.478~\log_{10}\left(\frac{M_*}{M_{\odot}}\right)+4.062 .
\end{equation}
We use the bisector fit because it is a monotonically increasing function of galaxy mass. The large differences at higher masses between the fits are largely due to the inclusion or exclusion of just four high-mass galaxies \citep{savaglio2005gemini}, illustrating the uncertainty at the extreme ends of \ac{MZ} relations. \citet{langer2006collapsar} approximate this fit with a simplified \ac{MZ} relation:
\begin{equation}
\label{eq:NL_mass_metallicity}
\frac{ M_*}{M_{x}} = \left(\frac{Z}{Z_{\odot}}\right)^2, 
\end{equation}
where $M_{\rm{x}}=7.64\times 10^{10}$\,M$_\odot$ \citep{panter2004mass}. 
\citet{langer2006collapsar} assume that the mean metallicity decreases exponentially with redshift as,
\begin{equation}
\label{eq:scaling}
\langle Z \rangle = Z_{\odot}10^{-0.3z}.
\end{equation}
When we translate this back into a \ac{MZ} relation we find that there is difference between the approximate \citet{langer2006collapsar} \ac{MZ} relation and the fit of \citet{savaglio2005gemini} (see Fig.~\ref{fig:MZrelations}). We introduce an offset to the model of \cite{langer2006collapsar} in order to  recover the relation by \citet{savaglio2005gemini}. This offset together with the original redshift scaling results in a high mean metallicity at redshift zero, but we keep this as an alternative model to look at its effects. 

The second \ac{MZ}-relation we consider is a theoretical model due to \citet{ma2015origin}. They combine cosmological simulations with stellar population synthesis models and a variety of feedback mechanisms to trace the evolution of the interstellar gas, for galaxy stellar masses ranging between $4 \leq \log_{10}(M_*/\mathrm{M}_\odot) \leq 11$ and redshifts between 0--6.  \citet{ma2015origin} give the \ac{MZ}-relationship as
\begin{equation}
\begin{split}
\label{eq:Ma2015}
\log_{10} \left( \frac{Z_\mathrm{gas}}{Z_\odot} \right) &= 0.35  \left[ \log_{10} \left( \frac{M_*}{M_{\odot}} \right)  - 10 \right]  \\
&  + 0.93  e^{-0.43 z} -1.05 .
\end{split}
\end{equation}
%

%%%%%%%%%%%%%%%%%%%%%%%%%%%%%%%%%%%%%%%%%%%%%%%%%%%%

\subsection{Galaxy Stellar Mass Density Function - GSMF}
\label{subsec:GSMF}

\begin{figure}
%\centering
%\begin{minipage}[t]{0.46\textwidth}
\includegraphics[width=\columnwidth]{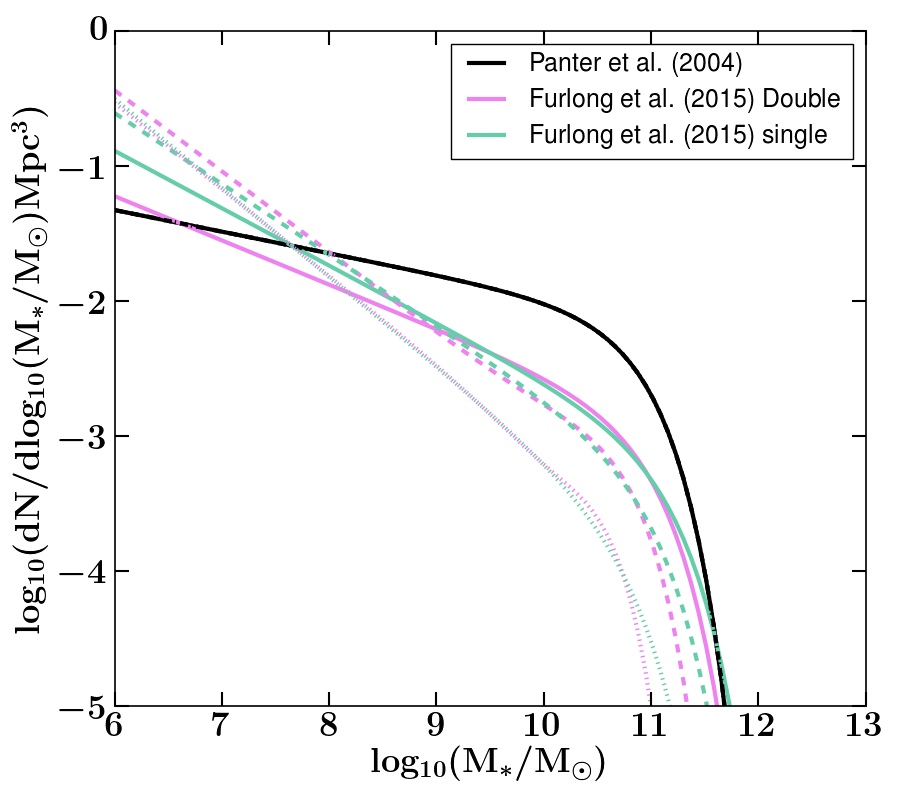}
\caption{Comparison between the galaxy stellar mass density functions at redshifts $z = 0$ (solid), $z = 1.5$ (dashed), and $z = 3$ (dotted). The relation by \citet{panter2004mass} is independent of redshift, therefore there is only a single curve (black). The double Schechter function by \citet{furlong2015evolution} (pink) has a steeper drop off at higher galaxy stellar masses compared to their single Schechter function (mint). Both power-law slopes of \citet{furlong2015evolution} are steeper than the Schechter function of \citet{panter2004mass}.  This shifts the distribution toward lower galaxy stellar masses, which translate to lower metallicities.}
\label{fig:GSMFrelations}
%\end{minipage}
\hfill
\end{figure}

The \ac{GSMF} is empirically constructed by converting the luminosity of a sample of galaxies into a stellar mass, assuming a mass-to-light ratio.  Although samples and methods differ between compilations, \citet{baldry2008galaxy} show that for galaxies within the mass range of $8.5 \leq \log_{10}(M_{*} / \mathrm{M}_\odot) \leq 12$ at redshift $z < 0.1$, there is good agreement on the shape of the \ac{GSMF}.  

The general shape is that of a Schechter function \citep{schechter1976analytic}:
\begin{equation}
\label{eq:Schechter}
\Phi_{M_*}(z) dM = \phi_1(z) \left( \frac{M_*}{M_c(z)} \right) ^{-\alpha(z)}
e^{-\frac{M_*}{M_c(z)}} dM, 
\end{equation} 
where $\alpha$ determines the slope of the \ac{GSMF} at the low-mass end, $M_c$ is the turnover mass, and $\phi_1$ the overall normalisation. However, a double Schechter function appears to better fit the extreme mass ends of the \ac{GSMF} \citep{baldry2008galaxy, furlong2015evolution}:
\begin{equation}
\begin{split}
\label{eq:doubleSchechter}
&\Phi_{M_*}(z) dM = e^{-\frac{M_*}{M_c(z)}} \times \\
& \left[ \phi_1(z) \left( \frac{M_*}{M_c(z)} \right) ^{-\alpha_1(z)} +
			\phi_2(z) \left( \frac{M_*}{M_c(z)} \right) ^{-\alpha_2(z)}
 \right]  dM.
\end{split}
\end{equation}

The double Schechter function fit determined from the EAGLE simulations by \citet{furlong2015evolution} is able to reproduce the empirical observations of \citet{duncan2014mass}. We performed a linear fit to the tabulated coefficient values in the appendix of \citet{furlong2015evolution} (see their table A1) to recover both a single and double Schechter \ac{GSMF}.  Their results are for redshifts in the range $0.1 < z < 4$ and we linearly interpolate the coefficients within that range.  We also extrapolate for lower and higher redshifts.  In order to avoid unphysical behaviour, we set $\phi_2$, which is zero at $z=0.1$ to also be zero at all redshifts below 0.1; fix $\alpha_2=-1.79$ at $z \leq 0.5$; and enforce $\alpha \geq -1.99$ everywhere.  This allows us to extrapolate the \citet{furlong2015evolution}  \ac{GSMF} over the full range $z \in [0,6.5]$.  

The \acp{GSMF} has an overall normalisation which in principle carries information on the star formation history, although \cite{furlong2015evolution} note that the normalisation of their fits is imperfect at the highest redshifts, while the slope remains well fitted.  However, we use a simplified model in which the \ac{SFR} is independent of the \ac{GSMF}, allowing us to independently parametrise and test the \ac{SFR} and the metallicity distribution.  Consequently, the normalisation coefficients $\phi$ are relevant only for describing the ratio between the two Schechter functions in Eq.\ref{eq:doubleSchechter}.

Figure \ref{fig:GSMFrelations} shows the different \ac{GSMF} relations at a few redshifts.  For comparison we also use a \textit{redshift-independent} single Schechter function of \citet{panter2004mass} as used in \citet{langer2006collapsar} and our previous work \citep{2018JimInference}. 

Even though the \citet{panter2004mass} \ac{GSMF} is redshift-independent, the metallicity distribution still changes due to the redshift dependence in the \ac{MZ} relation.  Meanwhile, the \citet{furlong2015evolution} \ac{GSMF} is redshift-dependent: as galaxies grow over time, the mass distribution shifts toward higher masses at lower redshifts \citep{duncan2014mass}.   Conversely, the masses are lower at higher redshifts, favouring lower metallicity.  Coupled with a redshift-dependent \ac{MZ} relation, this further reduces mean metallicity at higher redshifts. 

%%%%%%%%%%%%%%%%%%%%%%%%%%%%%%%%%%%%%%%%%

\subsection{Metallicity Specific Star Formation Rate - MSSFR}
\label{subsec:MSSFR}

The \ac{MZ} relation allows us to convert the \ac{GSMF} into a metallicity distribution $dP/dZ$ (the last term of Eq.~\ref{eq:MSSFR}).  In practice, when integrating over metallicity, we sum over discrete bins.  We convert the edges of those bins into limits on galaxy stellar masses in order to determine the fraction of star formation that happens in a given metallicity bin as the fraction of the \ac{GSMF} that falls into the appropriate mass range at a given redshift.

We convert the number density of Eq.~\ref{eq:doubleSchechter} into a mass density by multiplying by $M_*$.  The form of this equation makes it possible to carry out the mass integral analytically, with the amount of mass at $M_* \leq M_x$ given through the incomplete gamma functions $\hat{\Gamma}$:
\begin{equation}{\label{eq:incompleteGamma}}
\int_0^{M_x}M_*\Phi_{M_*} dM_* = \Phi_1 \hat{\Gamma}(\alpha_1+2, \frac{M_x}{M_c}) +
		      \Phi_2\hat{\Gamma}(\alpha_2+2, \frac{M_x}{M_c})
\end{equation}
The fraction of mass in the range between $M_x \leq M_* \leq M_y$ can be obtained from the above equation after normalisation with the complete gamma function $\Gamma$.  
Figure \ref{fig:ZvsPDF} shows several of the resulting star formation metallicity distributions at a few redshifts. %The mean metallicity \citet{langer2006collapsar} already favours lower metallicities at zero redshifts and furthermore evolves quicker towards even lower metallicities at higher redshifts. Note that the distribution by \citet{ma2015origin} appears highly extra solar because we use the definition of solar oxygen to hydrogen number density of 8.69 instead of their value of 9. 

We compute the \ac{MSSFR} by multiplying the metallicity distribution at a given redshift by the \ac{SFR} at that redshift (Eq.\ref{eq:MSSFR}).  Altogether we test the effect of 18 variations (2 \ac{SFR} $\times$ 3 \ac{MZ} $\times$ 3 \ac{GSMF}), as well as our preferred \ac{MSSFR} model. The two \ac{SFR} variations differ mostly at redshifts above 2. The \ac{MZ} relations span the range between extra-solar and sub-solar metallicities at $z=0$. The \acp{GSMF} variants include a static redshift-independent fit and two redshift-dependent fits, which evolve toward higher galaxy stellar masses at lower redshifts (see tables \ref{allRates}, \ref{allLikelihoods}). 

%%%%%%%%%%

\subsection{Definition of Solar Values}
\label{subsec:solar_values}

\begin{figure}
\includegraphics[width=\columnwidth]{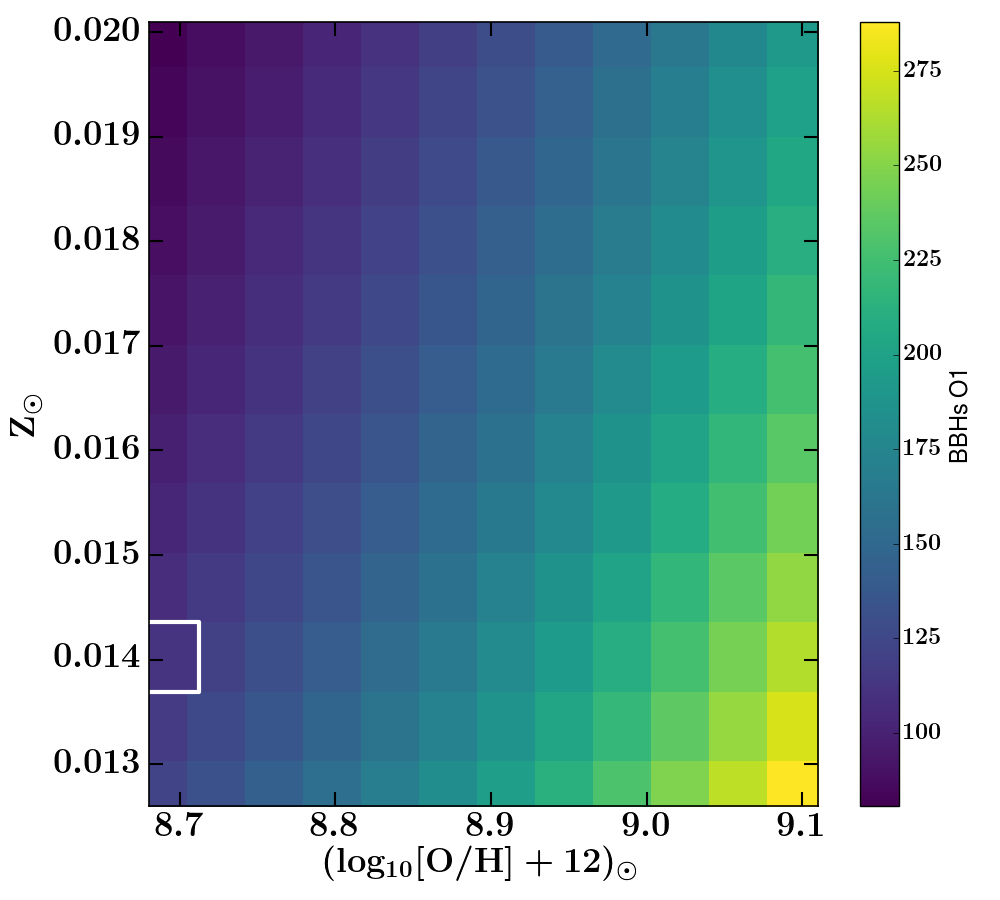}
      \caption{The predicted number of \ac{BBH} detections per year at the sensitivity of the  first observing run for different choices of the solar mass fraction $\rm Z_{\odot}$ and oxygen abundance $(\log_{10}[\mathrm{O/H}] + 12)_{\odot}$. The \ac{MSSFR} model is based on the \ac{SFR} of \citet{madau2014cosmic}, with the \ac{MZ} relation of \citet{ma2015origin} and the double Schechter \ac{GSMF} of \citet{furlong2015evolution}.  The same (pessimistic) evolutionary model is assumed. The white square denotes the point in parameter space we assumed for this study.}
    \label{fig:solarValues}
\end{figure}

In this study we defined the solar metallicity mass fraction as $ Z_{\odot}=0.0142$ and the solar oxygen abundance as $ \log_{10}[\mathrm{O/H}]_{\odot} + 12 = 8.69$ based on \citet{2009AsplundMetallicity}. However, the assumed solar values differ between papers, so our choice is not always consistent with the fits used.  

In particular, \citet{ma2015origin} assume a mass fraction of $Z_{\odot} = 0.02$ and a specific iron mass fraction of 0.00173 to obtain an oxygen abundance of $\log_{10}[\mathrm{O/H}]_{\odot} + 12 = 9.0$.  \citet{savaglio2005gemini} assume an oxygen abundance of 8.69, but mention that systematics can lead to uncertainties in the range between 8.7 and 9.1.  Meanwhile, their single stellar models for their galaxy models assume a mass fraction of $Z_{\odot} = 0.02$ \citep{1999LeithererStarBurstZ}.  On the other hand, \citet{furlong2015evolution} use a solar mass fraction of $Z_{\odot} = 0.0127$.  

We evaluate the impact of the assumed solar metallicity and oxygen abundances on our predictions by varying these within a single \ac{MSSFR} model.  Figure \ref{fig:solarValues}) shows the predicted number of \ac{BBH} detections per year at the sensitivity of the first observing run over a two-dimensional grid of solar metallicities and oxygen abundances, while keeping all other model parameters fixed.  We see that these uncertainties alone could change the predicted values by up to a factor of 3.  

%%%%%%%%%%%%%%%%%

\section{Remnant Masses of Single Stars}
\label{AppRemnant}

\begin{figure}
% \hspace{-20pt}
\includegraphics[width=\columnwidth]{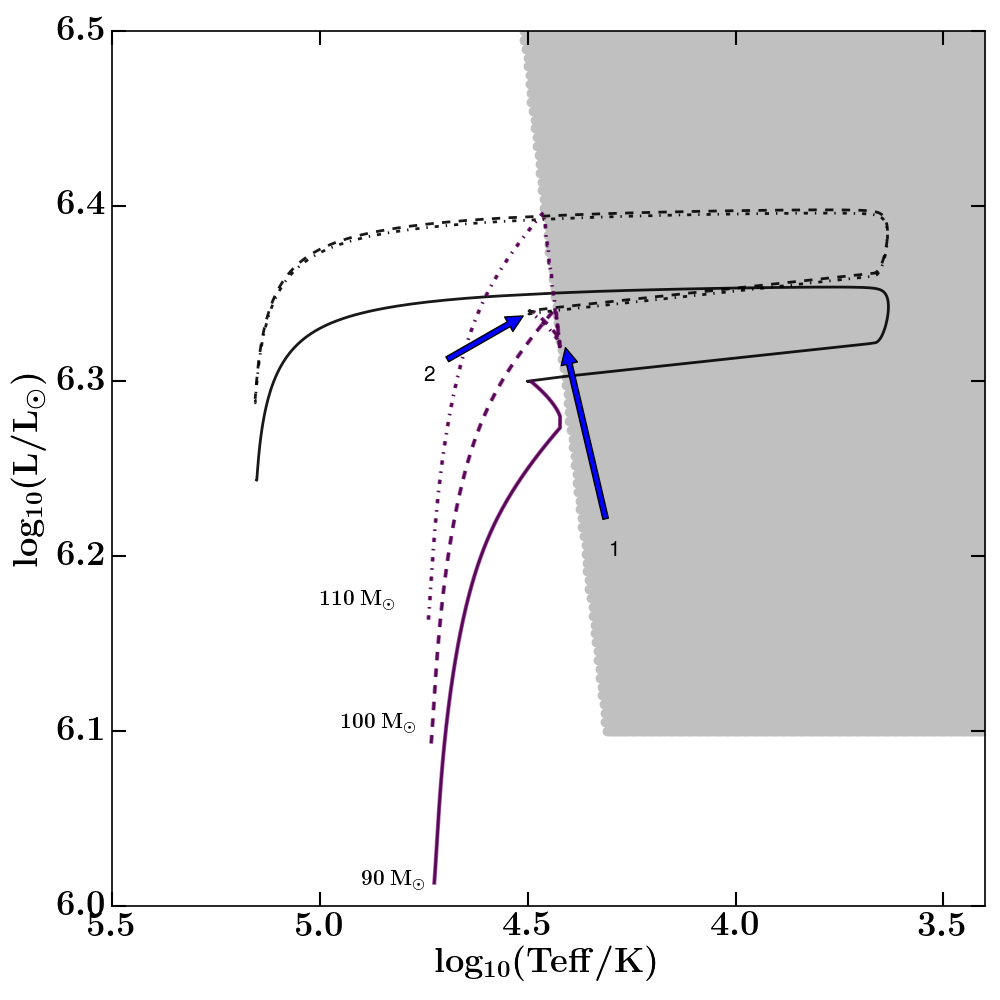}
     \caption{Single stellar tracks of stars with initial masses of 90~$\rm M_{\odot}$, 100~$\rm M_{\odot}$, and 110~$\rm M_{\odot}$. The metallicity of the stars is $Z = Z_{\odot}/3$. The purple portions of the tracks indicate the core hydrogen burning phase (main sequence), while subsequent evolution is shown in black. The region in which we apply the LBV-wind mass loss rate is shaded. At point 1 the 100~$\rm M_{\odot}$ and 110~$\rm M_{\odot}$ stars turn off the main sequence. At point 2 they start evolving onto the Hertzsprung gap. The two tracks evolve identically from point 1, resulting in the same remnant mass.}
     \label{fig:HR_LBV}
\end{figure}

% \begin{landscape}

% \end{landscape}

In COMPAS models, metallicity impacts the masses of compact remnants by influencing stellar evolutionary tracks and the rates of wind-driven mass loss.  The \citet{2012FryerRemnants} recipes for  calculating the remnant mass are based on the mass of the carbon-oxygen core and the total stellar mass at the moment of the supernova. If the carbon-oxygen core mass exceeds 16~$\rm M_{\odot}$, there is assumed to be no explosion and the remnant mass is the same as the total mass of the star before the supernova.  

However, within our models, all single stars above a certain initial mass yield the same remnant mass at a given metallicity.\footnote{These simulations do not include \acp{PISN} or pulsational \acp{PISN}. Future COMPAS analyses will incorporate them \citep{2019arXiv190402821S}.}  This is driven by our implementation of \ac{LBV}-wind mass loss (see Sec.~\ref{popsynth}). 

Figure~\ref{fig:HR_LBV} shows three tracks of very massive single stars. The shaded region is where we apply the \ac{LBV}-wind mass loss rates \citep{belczynski2010maximum}. When stars are on the main sequence (core hydrogen burning phase), they evolve on a nuclear timescale, which is not sufficiently fast to overcome the LBV winds and pass through the Humphreys-Davidson limit \citep{1994HumphreysLBV} into the shaded region. At point 1 the stars start to turn off the main sequence. By this time stars with initial masses of 100 and 110 solar masses have the same mass.
At point 2 they begin to evolve onto the Hertzsprung gap. It is at this point that the analytical fits of \citet{hurley2000comprehensive} define a core mass. This core mass only depends on the current properties of the stars, so the two stars continue evolving identically. Their faster, thermal-timescale evolution now allows them to pass through the Humphreys-Davidson limit  and enter the shaded region. They end up having the same remnant mass.

The process described above can yield sharp peaks in the \ac{BBH} mass distribution.  Every binary in which both stars go through this \ac{LBV} phase on the main sequence will end up with the same total \ac{BBH} mass. This is the maximum total mass for a given metallicity.
The lower the initial mass for this \ac{LBV}-wind mass loss, i.e., the higher the metallicity, the more binaries will have degenerate remnant masses.
Hence the sharpest peaks in the total \ac{BBH} mass distribution are for metallicities around a third solar 
(see Fig.~\ref{fig:Mtot}). This feature of the COMPAS and StarTrack implementation of \ac{LBV} winds also explains the asymptote of the maximal remnant mass in figure 1 of \citet{belczynski2010maximum} and the peaks in the highest mass bins of \citet{2012DominikCEE}.

%%%%%%%%%%%%%%%%%%

\section{Statistics}
\label{Bayes}

In this appendix, we describe our procedures for computing the likelihood of a given \ac{MSSFR} model given the observed number of detections and their chirp masses, and describe the use of bootstrapping to estimate the Monte Carlo simulation uncertainty.

\subsection{Evaluating model likelihoods}

We can write the total likelihood $\mathcal{L}_\mathrm{tot}(d|M)$ of observing the data set $d$, which consists of $N_\mathrm{obs}$ detections with individual data $d_i$, given a model $M$ that predicts $N_M$ expected detections with a probability distribution of source properties $P_M$, as \citep[e.g.,][]{Mandel:2018select}
\begin{equation}{\label{eq:Likely}}
\mathcal{L}_\mathrm{tot}(d|M) = \frac{N_M^{N_{obs}}}{N_{obs}!}e^{-N_M} \prod_{i=1}^{N_{obs}} p(d_i | P_M) .
\end{equation}
Here, we focus on the chirp mass $\mathcal{M}_c$, as the parameter which is best constrained by gravitational wave observations and is directly predicted by COMPAS simulations.   Writing the preceding equation in logarithmic form, the log-likelihood of a particular \ac{MSSFR} model is
\begin{equation}
\begin{split}
  \log_{10}\left(\mathcal{L}_\mathrm{tot}(d|M)\right)  ={}& \log_{10}\left(\mathcal{L}(N_\mathrm{obs}|N_M)\right)+ \\
  & \sum_{i = 1}^{N_\mathrm{obs}}\log_{10}\left(\mathcal{L}(\mathcal{M}_{c,i}|p_M(\mathcal{M}_{c}))\right),
\end{split}
\end{equation}
where $\mathcal{M}_{c,i}$ is the measured chirp-mass of the $i$'th gravitational wave observation and $p_M(\mathcal{M}_{c})$ is chirp-mass distribution characterising the \ac{MSSFR} model $M$.  The first term is abbreviated as $\mathcal{L}_\mathrm{R}$ in table \ref{allLikelihoods}.  The second term, $\mathcal{L}_\mathrm{M_{chirp}} \equiv \mathcal{L}(\mathcal{M}_{c,i}|p_M(\mathcal{M}_{c}))$ is the probability
of detecting a chirp mass $\mathcal{M}_{c,i}$ given the chirp-mass distribution predicted from the \ac{MSSFR} model $M$.

COMPAS Monte Carlo simulations yield a discrete set of chirp masses and their respective rates.
A kernel density estimator is used to turn this set of discrete data points into an approximated continuous function. 
We do this by approximating each of $N_\mathrm{sim}$ chirp masses produced by the COMPAS simulation as a 1-dimensional Gaussian centred on the simulated chirp-mass value $\mathcal{M}_j$.  All Gaussians have the same bandwidth $\sigma$, determined using the default `Scott's rule' \citep{scott2015multivariate} of the Gaussian kernel density estimator in the \texttt{scipy} package \citep{oliphant2007python, perez2011python}.
Each simulated data point $j$ contributes to the overall probability density function proportionally to its observing rate $R_{j}$, estimated in Eq.~\ref{eq:detections}.
Therefore we re-weigh each data point by $R_j$ and normalise by the total rate $R_\mathrm{tot}$,
\begin{equation}
\label{eq:KDE}
p_M(\mathcal{M}_{c})= \frac{1}{R_\mathrm{tot}} \sum_{j=1}^{N_\mathrm{sim}} R_{j} \frac{1}{\sqrt{2 \pi \sigma^2}} e^{-\frac{(\mathcal{M}_c-\mathcal{M}_j)^2}{2\sigma^2}} .
\end{equation}

\begin{figure}
\includegraphics[width=\columnwidth]{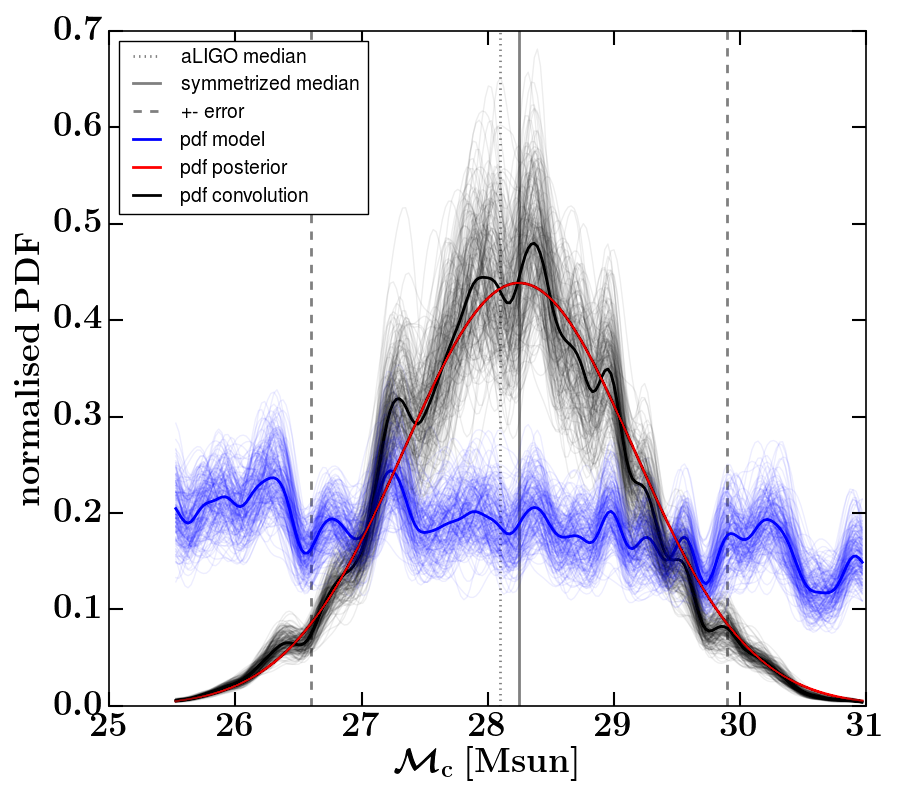}
      \caption{An example of the likelihood calculation for GW150914. The dashed vertical lines show the 90 per cent confidence
      interval from \citet{abbott2016binary}. The dotted vertical line is the median from  \citet{abbott2016binary} and the solid vertical line
      is the median after symmetrising. The red curve is a mock gaussian posterior. The blue curve is part of the normalised chirp mass
      distribution obtained by applying a one dimensional KDE to the results of the COMPAS simulation. The black curve shows the Gaussian likelihood convolved with the model. The fainter lines show scatter in the chirp-mass distribution and
      the convolution as estimated with bootstrapping.}\label{fig:Bayes}
\end{figure}

For a single perfect detection, the likelihood of observing a chirp-mass $\mathcal{M}_c$ would be given by Eq.~\ref{eq:KDE}.  In practice, gravitational-wave measurements suffer from observational uncertainty, although these are typically small for chirp masses.  Chirp-mass posteriors of individual detections were not yet available when this work started; therefore, we reconstruct them as symmetric Gaussian distributions with 90 per cent confidence intervals matching those reported in \citet{abbott2016binary, LVC2018O2}.  The reported error-bars are asymmetric,
so the median of our reconstructed posterior is slightly shifted compared to the original.  Given the accuracy of chirp mass measurement, we make two further simplifications.  We ignore the impact of the priors used in \citet{LVC2018O2} (which is reasonable inasmuch as the posterior is determined by the sharply peaked likelihood function), and do not reweigh by those priors; and we ignore the selection effects on the chirp mass for the purpose of population analysis, since the selection function does not vary significantly over the range of likelihood support (see \citet{Mandel:2018select} for a discussion of both issues).  With these simplifications, the likelihood of observing a particular gravitational wave event $i$, characterised by the approximated Gaussian posterior of the chirp mass $p_i(\mathcal{M}_c)$, given a MSSFR model $M$, becomes
\begin{equation}
 \mathcal{L}(\mathcal{M}_{c,i}|p_M(\mathcal{M}_c)) = \int_{0}^{\infty} p_i(\mathcal{M}_c)\ p_M(\mathcal{M}_c)\ d\mathcal{M}_c .
\end{equation}
Figure \ref{fig:Bayes} shows our constructed posterior for GW150914 (red); part of the chirp-mass distribution estimated from the \ac{MSSFR} model which combines the \ac{SFR} of \citet{madau2014cosmic}, the \ac{MZ} relation of \citet{ma2015origin} and the \ac{GSMF} of \citet{furlong2015evolution} (blue); and the 
convolution between the two (black). 
The integral of this convolution is our estimate of the likelihood  $\mathcal{L}(\mathcal{M}_{c,i}|p_M(\mathcal{M}_c))$.

\subsection{Bootstrapping}
Our simulation is based on a Monte Carlo sampling of binaries.  
%This means our results have an uncertainty due to sampling. 
%Ideally, to recover this uncertainty, we should apply our entire analysis, from binary evolution to post-processing and likelihood evaluations, to  many other sets of binaries, drawn from our initial parameter distributions. However, this is computationally unfeasible. 
We estimate the sampling uncertainty on all derived quantities via bootstrapping:
we uniformly re-sample a set with the same total number of binaries from our already evolved initial set of binaries (with replacement), including systems which did not form a \ac{DCO}. 
The central value in tables \ref{allRates}, \ref{allLikelihoods} corresponds to the original sample, while the error bars correspond to the 5th and 95th percentile rates and likelihoods from bootstrapping.

\begin{table*}
\begin{tabular}{|l |l | l| llllll}
\multicolumn{3}{|c|}{MSSFR Variation } & \multicolumn{2}{|c|}{BBH Rates} & \multicolumn{2}{|c|}{BHNS Rates} & \multicolumn{2}{|c|}{BNS Rates} \\
SFR & MZ & GSMF & $z=0$ merg. &  O1 det.   & $z=0$ merg. &  O1 det. & $z=0$ merg. &  O1 det. \\
& & &   Gpc$^{-3}$ yr$^{-1}$  & yr$^{-1}$ &   Gpc$^{-3}$ yr$^{-1}$ & yr$^{-1}$  & Gpc$^{-3}$ yr$^{-1}$ & yr$^{-1}$\\ \hline \hline
 &&&&&&&&\\
\multicolumn{9}{|c|}{Pessimistic} \\
 &&&&&&&&\\\hline
\multicolumn{3}{|c|}{Preferred model}  &$ 49.00_{-1.68}^{+1.93} $ & $ 21.80_{-0.50}^{+0.47} $ & $ 56.87_{-1.89}^{+1.80} $ & $ 0.07_{-0.00}^{+0.00} $ &  $ 20.00_{-1.03}^{+1.34} $ & $ 0.23_{-0.015}^{+0.02} $ \\ \hline
Madau et al. & Ma et al. (2004) & 1 &$ 63.07_{-1.76}^{+1.86} $ & $ 18.43_{-0.4}^{+0.42} $ & $ 32.22_{-1.43}^{+1.7} $ & $ 0.4_{-0.02}^{+0.02} $ & $ 85.97_{-2.41}^{+2.47}$ & $ 0.11_{-0.0}^{+0.0} $ \\
 & & 2 &$ 158.56_{-2.48}^{+2.07} $ & $ 94.35_{-1.39}^{+1.37} $ & $ 40.73_{-1.13}^{+1.37} $ & $ 0.51_{-0.02}^{+0.02} $ & $ 90.87_{-2.15}^{+2.39}$ & $ 0.12_{-0.0}^{+0.0} $ \\
 & & 3 &$ 174.71_{-2.7}^{+2.28} $ & $ 113.92_{-1.22}^{+1.12} $ & $ 42.14_{-1.22}^{+1.43} $ & $ 0.52_{-0.02}^{+0.02} $ & $ 91.42_{-2.24}^{+2.37}$ & $ 0.13_{-0.0}^{+0.0} $ \\\cline{2-9}
 & Langer et al.  &1 &$ 448.84_{-5.24}^{+4.28} $ & $ 247.22_{-2.78}^{+2.63} $ & $ 95.47_{-2.12}^{+1.85} $ & $ 1.28_{-0.03}^{+0.03} $ & $ 144.57_{-2.3}^{+2.08}$ & $ 0.22_{-0.0}^{+0.0} $ \\
 &  &2 &$ 563.44_{-5.9}^{+4.1} $ & $ 441.08_{-6.24}^{+5.72} $ & $ 91.76_{-1.8}^{+1.73} $ & $ 1.19_{-0.03}^{+0.03} $ & $ 143.81_{-2.1}^{+1.93}$ & $ 0.22_{-0.0}^{+0.0} $ \\
 &  &3 &$ 589.13_{-6.05}^{+4.27} $ & $ 492.27_{-6.63}^{+5.85} $ & $ 96.32_{-1.9}^{+1.9} $ & $ 1.25_{-0.03}^{+0.03} $ & $ 147.1_{-2.3}^{+1.92}$ & $ 0.23_{-0.0}^{+0.0} $ \\\cline{2-9}
 & Langer et al., offset  &1 &$ 59.17_{-1.34}^{+1.27} $ & $ 28.72_{-0.45}^{+0.46} $ & $ 18.98_{-0.77}^{+1.01} $ & $ 0.23_{-0.01}^{+0.01} $ & $ 72.09_{-3.01}^{+3.21}$ & $ 0.09_{-0.0}^{+0.0} $ \\
 &  &2 &$ 151.84_{-2.23}^{+1.88} $ & $ 120.3_{-2.23}^{+2.12} $ & $ 28.67_{-0.88}^{+0.92} $ & $ 0.35_{-0.01}^{+0.01} $ & $ 79.27_{-2.52}^{+2.67}$ & $ 0.11_{-0.0}^{+0.0} $ \\
 &  &3 &$ 167.64_{-2.29}^{+1.92} $ & $ 148.74_{-2.05}^{+2.0} $ & $ 29.4_{-0.98}^{+0.97} $ & $ 0.35_{-0.01}^{+0.01} $ & $ 79.32_{-2.51}^{+2.57}$ & $ 0.11_{-0.0}^{+0.0} $ \\\hline
Strolger et al. & Ma et al. (2004) & 1 &$ 101.98_{-2.88}^{+3.01} $ & $ 32.93_{-0.84}^{+0.83} $ & $ 42.11_{-1.81}^{+2.4} $ & $ 0.52_{-0.02}^{+0.03} $ & $ 91.66_{-2.49}^{+2.71}$ & $ 0.12_{-0.0}^{+0.0} $ \\
 & &2 &$ 255.46_{-5.7}^{+4.49} $ & $ 203.93_{-4.76}^{+5.17} $ & $ 49.05_{-1.68}^{+2.1} $ & $ 0.6_{-0.02}^{+0.02} $ & $ 98.87_{-2.72}^{+2.59}$ & $ 0.14_{-0.0}^{+0.0} $ \\
 & &3 &$ 271.91_{-4.67}^{+4.09} $ & $ 208.21_{-3.38}^{+3.29} $ & $ 50.49_{-1.66}^{+1.98} $ & $ 0.61_{-0.02}^{+0.03} $ & $ 98.72_{-2.72}^{+2.4}$ & $ 0.14_{-0.0}^{+0.0} $ \\\cline{2-9}
 & Langer et al. &1 &$ 574.91_{-8.19}^{+6.58} $ & $ 406.39_{-5.66}^{+5.25} $ & $ 99.35_{-2.43}^{+2.32} $ & $ 1.28_{-0.04}^{+0.03} $ & $ 153.78_{-2.9}^{+2.74}$ & $ 0.23_{-0.0}^{+0.0} $ \\
 &  &2 &$ 688.91_{-10.42}^{+9.03} $ & $ 659.25_{-14.58}^{+15.27} $ & $ 95.9_{-2.62}^{+2.65} $ & $ 1.19_{-0.03}^{+0.03} $ & $ 153.93_{-3.4}^{+3.49}$ & $ 0.24_{-0.01}^{+0.01} $ \\
 &  &3 &$ 714.15_{-11.17}^{+8.77} $ & $ 710.91_{-14.29}^{+15.17} $ & $ 100.12_{-2.69}^{+2.67} $ & $ 1.25_{-0.03}^{+0.03} $ & $ 157.4_{-3.43}^{+3.8}$ & $ 0.24_{-0.01}^{+0.01} $ \\\cline{2-9}
 & Langer et al., offset &1 &$ 132.55_{-3.54}^{+3.38} $ & $ 89.79_{-1.42}^{+1.42} $ & $ 27.88_{-1.27}^{+1.96} $ & $ 0.33_{-0.01}^{+0.02} $ & $ 79.81_{-3.29}^{+3.16}$ & $ 0.11_{-0.0}^{+0.0} $ \\
 &  &2 &$ 259.14_{-6.53}^{+5.65} $ & $ 267.34_{-7.55}^{+8.2} $ & $ 36.67_{-1.57}^{+1.58} $ & $ 0.43_{-0.02}^{+0.02} $ & $ 88.47_{-3.58}^{+3.1}$ & $ 0.12_{-0.0}^{+0.01} $ \\
 &  &3 &$ 276.57_{-6.48}^{+5.32} $ & $ 292.76_{-6.55}^{+7.71} $ & $ 37.36_{-1.56}^{+1.64} $ & $ 0.43_{-0.02}^{+0.02} $ & $ 88.34_{-3.29}^{+2.88}$ & $ 0.12_{-0.0}^{+0.0} $ \\
\hline 
 &&&&&&&&\\
\multicolumn{9}{|c|}{Optimistic} \\
 &&&&&&&&\\\hline
\multicolumn{3}{|c|}{Preferred model}  &$ 190.85_{-3.99}^{+3.94} $ & $ 36.80_{-0.65}^{+0.65} $ & $ 158.8_{-3.49}^{+3.13} $  & $ 0.23_{-0.01}^{+0.00} $  & $ 56.68_{-1.87}{+2.21} $ & $ 0.51_{-0.02}^{+0.02} $ \\ \hline
Madau et al. & Ma et al. (2004) & 1 &$ 291.23_{-5.28}^{+4.93} $ & $ 46.27_{-0.72}^{+0.7} $ & $ 89.32_{-2.47}^{+3.19} $ & $ 0.85_{-0.02}^{+0.03} $ & $ 231.08_{-4.8}^{+4.36}$ & $ 0.33_{-0.01}^{+0.01} $  \\
 & & 2 &$ 408.74_{-5.52}^{+4.03} $ & $ 128.37_{-1.48}^{+1.44} $ & $ 102.22_{-2.07}^{+2.67} $ & $ 1.04_{-0.02}^{+0.03} $ & $ 227.38_{-4.46}^{+4.13}$ & $ 0.33_{-0.01}^{+0.01} $  \\
 & & 3 &$ 431.55_{-5.58}^{+4.5} $ & $ 148.55_{-1.38}^{+1.26} $ & $ 104.83_{-2.18}^{+2.71} $ & $ 1.06_{-0.02}^{+0.03} $ & $ 226.34_{-4.21}^{+3.71}$ & $ 0.33_{-0.01}^{+0.01} $  \\\cline{2-9}
 & Langer et al.  &1 &$ 938.86_{-7.01}^{+6.16} $ & $ 332.4_{-2.77}^{+2.92} $ & $ 201.14_{-3.1}^{+3.44} $ & $ 2.33_{-0.04}^{+0.04} $ & $ 209.7_{-2.73}^{+2.85}$ & $ 0.33_{-0.0}^{+0.0} $  \\
 &  &2 &$ 1002.87_{-6.55}^{+5.33} $ & $ 517.59_{-6.39}^{+5.41} $ & $ 187.17_{-2.49}^{+2.79} $ & $ 2.16_{-0.03}^{+0.04} $ & $ 210.0_{-2.48}^{+2.47}$ & $ 0.33_{-0.0}^{+0.0} $  \\
 &  &3 &$ 1042.93_{-6.97}^{+5.27} $ & $ 572.54_{-6.41}^{+5.69} $ & $ 193.78_{-2.65}^{+2.98} $ & $ 2.25_{-0.03}^{+0.04} $ & $ 208.21_{-2.49}^{+2.55}$ & $ 0.33_{-0.0}^{+0.0} $  \\\cline{2-9}
 & Langer et al., offset  &1 &$ 190.7_{-3.65}^{+3.5} $ & $ 43.44_{-0.52}^{+0.52} $ & $ 59.21_{-2.15}^{+2.69} $ & $ 0.52_{-0.01}^{+0.02} $ & $ 236.54_{-6.59}^{+6.48}$ & $ 0.34_{-0.01}^{+0.01} $  \\
 &  &2 &$ 317.64_{-4.02}^{+3.16} $ & $ 141.94_{-2.08}^{+2.28} $ & $ 74.72_{-1.91}^{+2.18} $ & $ 0.73_{-0.02}^{+0.02} $ & $ 232.11_{-5.96}^{+5.66}$ & $ 0.34_{-0.01}^{+0.01} $  \\
 &  &3 &$ 335.02_{-4.01}^{+3.3} $ & $ 170.2_{-2.02}^{+2.17} $ & $ 75.38_{-1.9}^{+2.2} $ & $ 0.73_{-0.02}^{+0.02} $ & $ 231.52_{-5.52}^{+5.68}$ & $ 0.33_{-0.01}^{+0.01} $  \\\hline
Strolger et al. & Ma et al. (2004) & 1 &$ 361.59_{-6.7}^{+5.79} $ & $ 64.2_{-1.12}^{+1.08} $ & $ 111.51_{-3.31}^{+4.29} $ & $ 1.09_{-0.03}^{+0.04} $ & $ 238.44_{-4.66}^{+4.5}$ & $ 0.34_{-0.01}^{+0.01} $  \\
 & &2 &$ 523.11_{-8.21}^{+5.83} $ & $ 239.42_{-4.94}^{+5.7} $ & $ 116.48_{-2.7}^{+3.27} $ & $ 1.18_{-0.03}^{+0.03} $ & $ 235.98_{-4.62}^{+4.6}$ & $ 0.34_{-0.01}^{+0.01} $  \\
 & &3 &$ 547.87_{-7.09}^{+6.08} $ & $ 244.58_{-3.41}^{+3.4} $ & $ 119.46_{-2.75}^{+3.05} $ & $ 1.22_{-0.03}^{+0.03} $ & $ 234.53_{-4.63}^{+3.89}$ & $ 0.34_{-0.01}^{+0.01} $  \\\cline{2-9}
 & Langer et al. &1 &$ 1061.19_{-9.88}^{+8.04} $ & $ 486.78_{-5.84}^{+5.82} $ & $ 204.37_{-3.35}^{+3.37} $ & $ 2.33_{-0.04}^{+0.05} $ & $ 220.3_{-3.03}^{+3.44}$ & $ 0.34_{-0.01}^{+0.01} $  \\
 &  &2 &$ 1118.71_{-10.66}^{+8.74} $ & $ 730.55_{-14.5}^{+14.82} $ & $ 189.09_{-3.06}^{+2.96} $ & $ 2.14_{-0.04}^{+0.04} $ & $ 221.11_{-3.54}^{+4.03}$ & $ 0.34_{-0.01}^{+0.01} $  \\
 &  &3 &$ 1157.22_{-11.19}^{+9.06} $ & $ 785.44_{-14.23}^{+14.83} $ & $ 194.98_{-3.31}^{+3.02} $ & $ 2.22_{-0.04}^{+0.04} $ & $ 219.59_{-3.66}^{+3.93}$ & $ 0.34_{-0.01}^{+0.01} $  \\\cline{2-9}
 & Langer et al., offset &1 &$ 291.8_{-5.44}^{+5.57} $ & $ 107.66_{-1.53}^{+1.37} $ & $ 76.78_{-2.67}^{+3.4} $ & $ 0.7_{-0.02}^{+0.02} $ & $ 244.96_{-6.62}^{+6.31}$ & $ 0.35_{-0.01}^{+0.01} $  \\
 &  &2 &$ 440.74_{-7.95}^{+6.82} $ & $ 290.5_{-7.21}^{+8.53} $ & $ 87.17_{-2.77}^{+2.78} $ & $ 0.86_{-0.02}^{+0.03} $ & $ 240.95_{-6.05}^{+6.21}$ & $ 0.35_{-0.01}^{+0.01} $  \\
 &  &3 &$ 460.27_{-7.84}^{+6.42} $ & $ 315.88_{-6.5}^{+7.94} $ & $ 87.82_{-2.54}^{+2.79} $ & $ 0.86_{-0.03}^{+0.02} $ & $ 240.19_{-5.99}^{+6.13}$ & $ 0.34_{-0.01}^{+0.01} $  \\
\end{tabular}
\caption{Table showing the merger and detection rates per \ac{DCO} type. The columns labeled `$z=0$ merg.' are the merger rate per year per cubic gigaparsec at zero redshift without selection effects; columns labeled `O1 det.' are the expected rate of detections per year at the sensitivity of the first observing run.  The error bars show the 90 per cent confidence interval due to Monte Carlo sampling evaluated via bootstrapping. The numbers in the column \ac{GSMF} refer to 1=\citet{panter2004mass}, 2=\citet{furlong2015evolution} (single Schechter function), 3=\citet{furlong2015evolution} (double Schechter function). Optimistic and pessimistic variants relate to the ability to eject the common envelope when the donor is a Hertzsprung-gap star. }\label{allRates}
\end{table*}

\begin{table*}
\begin{tabular}{|l |l | l| lll}
\multicolumn{3}{|c|}{MSSFR Variation} & \multicolumn{3}{|c|}{ Likelihoods ($\log_{10}$)}  \\
SFR & MZ & GSMF & $\mathcal{L}_\mathrm{M_{c}}$  &    $\mathcal{L}_\mathrm{R}$ & $\rm \mathcal{L}_{tot}$ \\ \hline \hline
 &&&&&\\
\multicolumn{6}{|c|}{Pessimistic} \\
 &&&&&\\ \hline
\multicolumn{3}{|c|}{Preferred model} &$ -32.32_{-0.18}^{+0.16} $ & $-0.90_{-0.00}^{+0.00}$  & $-33.22_{-0.18}^{+0.16}$ \\ \hline
Madau et al. & Ma et al. (2004) & 1 &$ -33.9_{-0.16}^{+0.14} $ & $ -0.97_{-0.02}^{+0.01} $ & $ -34.87_{-0.16}^{+0.14} $   \\
 & & 2 &$ -32.42_{-0.08}^{+0.07} $ & $ -8.86_{-0.21}^{+0.21} $ & $ -41.28_{-0.26}^{+0.24} $   \\
 & & 3 &$ -32.48_{-0.07}^{+0.07} $ & $ -11.9_{-0.18}^{+0.19} $ & $ -44.38_{-0.22}^{+0.23} $   \\\cline{2-6}
 & Langer et al.  &1 &$ -32.24_{-0.05}^{+0.05} $ & $ -34.85_{-0.47}^{+0.5} $ & $ -67.09_{-0.47}^{+0.49} $   \\
 &  &2 &$ -32.61_{-0.06}^{+0.06} $ & $ -70.6_{-1.05}^{+1.23} $ & $ -103.21_{-1.05}^{+1.22} $   \\
 &  &3 &$ -32.77_{-0.07}^{+0.06} $ & $ -80.23_{-1.04}^{+1.32} $ & $ -113.0_{-1.09}^{+1.3} $   \\\cline{2-6}
 & Langer et al., offset  &1 &$ -32.3_{-0.1}^{+0.09} $ & $ -1.07_{-0.02}^{+0.02} $ & $ -33.38_{-0.1}^{+0.09} $   \\
 &  &2 &$ -32.68_{-0.08}^{+0.08} $ & $ -12.93_{-0.33}^{+0.38} $ & $ -45.61_{-0.38}^{+0.42} $   \\
 &  &3 &$ -32.87_{-0.07}^{+0.07} $ & $ -17.62_{-0.33}^{+0.35} $ & $ -50.49_{-0.36}^{+0.38} $   \\\hline
Strolger et al. & Ma et al. (2004) & 1 &$ -33.82_{-0.17}^{+0.14} $ & $ -1.31_{-0.05}^{+0.06} $ & $ -35.13_{-0.17}^{+0.18} $   \\
 & &2 &$ -32.81_{-0.1}^{+0.11} $ & $ -27.14_{-0.91}^{+0.87} $ & $ -59.95_{-0.97}^{+0.93} $   \\
 & &3 &$ -32.65_{-0.08}^{+0.08} $ & $ -27.9_{-0.58}^{+0.61} $ & $ -60.54_{-0.61}^{+0.63} $   \\\cline{2-6}
 & Langer et al. &1 &$ -32.44_{-0.06}^{+0.06} $ & $ -64.11_{-0.97}^{+1.08} $ & $ -96.55_{-1.02}^{+1.08} $   \\
 &  &2 &$ -32.98_{-0.1}^{+0.1} $ & $ -111.92_{-2.81}^{+2.83} $ & $ -144.9_{-2.83}^{+2.9} $   \\
 &  &3 &$ -33.09_{-0.1}^{+0.09} $ & $ -121.79_{-2.83}^{+2.8} $ & $ -154.87_{-2.9}^{+2.86} $   \\\cline{2-6}
 & Langer et al., offset &1 &$ -32.46_{-0.1}^{+0.08} $ & $ -8.18_{-0.21}^{+0.21} $ & $ -40.63_{-0.26}^{+0.23} $   \\
 &  &2 &$ -33.2_{-0.12}^{+0.12} $ & $ -38.48_{-1.48}^{+1.44} $ & $ -71.68_{-1.51}^{+1.51} $   \\
 &  &3 &$ -33.22_{-0.11}^{+0.11} $ & $ -43.1_{-1.27}^{+1.29} $ & $ -76.33_{-1.34}^{+1.35} $   \\
\hline 
 &&&&&\\
\multicolumn{6}{|c|}{Optimistic} \\
 &&&&&\\ \hline
\multicolumn{3}{|c|}{Preferred model} & $-33.1_{-0.11}^{+0.08}$  & $-1.58_{-0.05}^{+0.05}$  & $-34.73_{-0.11}^{+0.12}$ \\ \hline
Madau et al. & Ma et al. (2004) & 1 &$ -36.37_{-0.1}^{+0.11} $ & $ -2.46_{-0.07}^{+0.07} $ & $ -38.84_{-0.14}^{+0.16} $   \\
 & & 2 &$ -32.61_{-0.06}^{+0.05} $ & $ -14.24_{-0.23}^{+0.25} $ & $ -46.85_{-0.26}^{+0.27} $   \\
 & & 3 &$ -32.52_{-0.05}^{+0.04} $ & $ -17.59_{-0.21}^{+0.24} $ & $ -50.1_{-0.22}^{+0.24} $   \\\cline{2-6}
 & Langer et al.  &1 &$ -32.55_{-0.04}^{+0.05} $ & $ -50.38_{-0.52}^{+0.52} $ & $ -82.93_{-0.53}^{+0.5} $   \\
 &  &2 &$ -32.55_{-0.05}^{+0.04} $ & $ -85.01_{-0.98}^{+1.21} $ & $ -117.56_{-1.02}^{+1.23} $   \\
 &  &3 &$ -32.67_{-0.05}^{+0.04} $ & $ -95.41_{-1.07}^{+1.28} $ & $ -128.08_{-1.11}^{+1.29} $   \\\cline{2-6}
 & Langer et al., offset  &1 &$ -32.85_{-0.06}^{+0.07} $ & $ -2.18_{-0.05}^{+0.05} $ & $ -35.03_{-0.08}^{+0.09} $   \\
 &  &2 &$ -32.49_{-0.06}^{+0.06} $ & $ -16.48_{-0.37}^{+0.39} $ & $ -48.97_{-0.39}^{+0.43} $   \\
 &  &3 &$ -32.59_{-0.06}^{+0.05} $ & $ -21.27_{-0.36}^{+0.35} $ & $ -53.85_{-0.36}^{+0.38} $   \\\hline
Strolger et al. & Ma et al. (2004) & 1 &$ -35.37_{-0.12}^{+0.11} $ & $ -4.58_{-0.14}^{+0.15} $ & $ -39.95_{-0.19}^{+0.21} $   \\
 & &2 &$ -32.54_{-0.08}^{+0.08} $ & $ -33.45_{-0.99}^{+0.89} $ & $ -65.99_{-0.98}^{+0.95} $   \\
 & &3 &$ -32.39_{-0.05}^{+0.05} $ & $ -34.37_{-0.6}^{+0.62} $ & $ -66.77_{-0.62}^{+0.65} $   \\\cline{2-6}
 & Langer et al. &1 &$ -32.38_{-0.05}^{+0.05} $ & $ -79.19_{-1.08}^{+1.11} $ & $ -111.57_{-1.11}^{+1.14} $   \\
 &  &2 &$ -32.78_{-0.09}^{+0.08} $ & $ -125.54_{-2.83}^{+2.79} $ & $ -158.32_{-2.86}^{+2.82} $   \\
 &  &3 &$ -32.86_{-0.08}^{+0.08} $ & $ -136.06_{-2.82}^{+2.78} $ & $ -168.92_{-2.88}^{+2.79} $   \\\cline{2-6}
 & Langer et al., offset &1 &$ -32.24_{-0.07}^{+0.06} $ & $ -10.91_{-0.21}^{+0.24} $ & $ -43.15_{-0.24}^{+0.23} $   \\
 &  &2 &$ -32.83_{-0.1}^{+0.1} $ & $ -42.69_{-1.51}^{+1.46} $ & $ -75.52_{-1.51}^{+1.53} $   \\
 &  &3 &$ -32.84_{-0.08}^{+0.09} $ & $ -47.34_{-1.35}^{+1.31} $ & $ -80.18_{-1.36}^{+1.31} $   \\
\end{tabular}
\caption{Table showing the log likelihoods of observing the rate and chirp mass distribution of \ac{BBH} mergers detected during the first two observing runs, within our default binary evolution model and for a range of \ac{MSSFR} variations.  $\rm \mathcal{L}_{tot}$ is the total likelihood,  $\rm \mathcal{L}_{R}$ is the Poisson likelihood of observing 10 \ac{BBH} events over 166 days of coincident observation, and  $\rm \mathcal{L}_{M_c}$ is the likelihood of observing the chirp-mass distribution. 
The error bars show the 90 per cent confidence interval due to Monte Carlo sampling evaluated via bootstrapping. 
The numbers in the column \ac{GSMF} refer to 1=\citet{panter2004mass}, 2=\citet{furlong2015evolution} (single Schechter function), 3=\citet{furlong2015evolution} (double Schechter function). Optimistic and pessimistic variants relate to the ability to eject the common envelope when the donor is a Hertzsprung-gap star.}\label{allLikelihoods}
\end{table*}

% Don't change these lines
\bsp	% typesetting comment
\label{lastpage}
\end{document}